\documentclass[preprint,12pt]{elsarticle}
\usepackage{xcolor}
\usepackage[export]{adjustbox}[2011/08/13]
\usepackage{caption}
\usepackage{subcaption}
\usepackage{multicol}
\usepackage{color, colortbl}
\usepackage{multirow}
\definecolor{LightBlue}{rgb}{0.9,0.9,0.9}

\usepackage{graphicx}
\usepackage{amssymb}

\usepackage{amsmath}

\usepackage{footnote}
\makesavenoteenv{tabular}
\makesavenoteenv{table}

\usepackage{siunitx}

\usepackage{hyperref}

\usepackage{cleveref}

\usepackage[symbol]{footmisc}

\usepackage{lineno}

\biboptions{sort&compress}

\journal{Computer Physics Communications}

\begin{document}

\begin{frontmatter}

\title{ADBSat: Verification and validation of a novel panel method for quick aerodynamic analysis of satellites}

\author[a]{Luciana A. Sinpetru\corref{cor}}
\author[a]{Nicholas H. Crisp}

\cortext[cor]{Corresponding author.\\\textit{E-mail address:} luciana.sinpetru@manchester.ac.uk}

\author[a]{Peter C. E. Roberts}
 \author[b]{Valeria Sulliotti-Linner}
 \author[c]{Virginia Hanessian}
 \author[d]{Georg H. Herdrich}
 \author[d]{Francesco Romano}
 \author[e]{Daniel Garcia-Almi\~nana}
 \author[e]{S\'ilvia Rodr\'iguez-Donaire}
 \author[f]{Simon Seminari}

\address[a]{The University of Manchester, Oxford Road, Manchester, M13 9PL, United Kingdom}
 \address[b]{Elecnor Deimos Satellite Systems, Calle Francia 9, 13500, Puertollano, Spain}
 \address[c]{GomSpace A/S, Langagervej 6, 9220 Aalborg East, Denmark}
 \address[d]{University of Stuttgart, Pfaffenwaldring 29, 70569 Stuttgart, Germany}
 \address[e]{UPC-BarcelonaTECH, Colom 11, TR5 - 08222 Terrassa, Barcelona, Spain}
 \address[f]{Euroconsult, 86 Boulevard de S\'ebastopol, 75003 Paris, France}

\begin{abstract}

We present the validation of ADBSat, a novel implementation of the panel method including a fast pseudo-shading algorithm, that can quickly and accurately determine the forces and torques on satellites in free-molecular flow. Our main method of validation is comparing test cases between ADBSat, the current de facto standard of direct simulation Monte Carlo (DSMC), and published literature. ADBSat exhibits a significantly shorter runtime than DSMC and performs well, except where deep concavities are present in the satellite models. The shading algorithm also experiences problems when a large proportion of the satellite surface area is oriented parallel to the flow, but this can be mitigated by examining the body at small angles to this configuration ($\pm$ \SI{0.1}{\degree}). We recommend that an error interval on ADBSat outputs of up to 3\% is adopted. Therefore, ADBSat is a suitable tool for quickly determining the aerodynamic characteristics of a wide range of satellite geometries in different environmental conditions in VLEO. It can also be used in a complementary manner to identify cases that warrant further investigation using other numerical-based methods.

\end{abstract}


\end{frontmatter}

\section{Introduction}
\label{S:intro}

Recent years have seen significant interest in the sustained operation of spacecraft at the bottom end of the low-earth orbit (LEO) range. Work has been focused on flight at orbital altitudes of \SIrange{100}{450}{\kilo\metre}, known as very low Earth orbits (VLEO) \cite{RobertsEtAl, DefinitionOfVLEO, CHAMP_firstpaper, GOCE_firstpaper, GRACE_firstpaper, GRACE_FO,SLATS, VLEOpropulsion, VLEOstabilization, VLEOionstuff, Crisp2020}. 
Operating a satellite in a low-altitude orbit can offer extensive benefits, in particular for Earth observation missions such as naval intelligence, fishing surveys, forestry management, and natural disaster support. Principally, the same payload orbiting closer to the Earth's surface will yield a better data resolution, with potentially more accurate positioning. Alternatively, a reduction in payload power will yield data of the same detail as a larger apparatus which orbits at a higher altitude. With this decrease in power requirements comes a desirable decrease in characteristics such as payload mass, size, and cost. All these factors together reduce the cost of manufacturing, launch and operation \cite{RobertsEtAl,Crisp2020,DefinitionOfVLEO}.

However, an important disadvantage in VLEO is the existence of atmospheric drag, which has a significant negative effect on spacecraft orbits. It leads to premature de-orbiting and a significant shortening of mission lifetime. The drag force can be quantified through the drag coefficient, $C_d$, which is invaluable to determining the drag response of a body in a fluid environment. ADBSat is a new program which determines the $C_d$ of any body, both quickly and accurately \cite{mySoftwarePaper, MostazaThesis}. It is a novel implementation of a panel method, in which the equations describing the interaction between the satellite surfaces and the atmospheric particles, known as the gas-surface interaction model (GSIM) equations, are used to calculate the desired outputs \cite{sabrina}. This program overcomes the difficulty of prohibitively complex GSIM formulae by treating the spacecraft as a set of flat triangular panels. Hence, it reduces the detailed spacecraft geometry to a set of simple shapes, to which the GSIM equations are easily applied. A summation of the plates' contributions provides the results for the body as a whole. A shading algorithm based on 2-dimensional projections of the spacecraft panels is also employed, with the aim of increasing accuracy for concave geometries \cite{mySoftwarePaper}. Past applications of the panel method such as DACFREE \citep{orion, apollo}, FreeMat \citep{FreeMat} and FreeMac \citep{FreeMac} and have suffered from a lack of reproducible and verifiable validation, and thus a limited knowledge of their accuracy.

A full, detailed description of the workings and implementation of ADBSat is available in an accompanying paper \cite{mySoftwarePaper}. This paper aims to verify and validate ADBSat, by comparing its results to those of two other common methods of determining $C_d$, direct simulation Monte Carlo (DSMC), and closed-form equations. Here, we shall only outline those simulation features which are critical to the verification and validation process. 

DSMC is a numerical technique that simulates the particles of the atmosphere by approximating many molecules as one simulation particle. First implemented by \citet{bird1994}, in recent times it has become `` ...the de facto method for modeling rarefied flow in the transition regime'' \citep{DragModelling}. It relies on embedding a computer-aided design (CAD) model of the spacecraft geometry in a computational domain, into which simulated particles are inserted. The drag force, which results from particles impinging on the shape, can then be measured. It accurately recreates the physics in this regime \citep{dsmcFoamPlus, dsmcFoam4, dsmcFoam2}. However, a high computational effort and time expense is required, due to the large number of particles in each simulation. As a result, its principal use to date for orbital aerodynamics has been in mission support \cite{ParkEtAlShapeOpt,intakeDesign} as well as analysis of the aerodynamic characteristics of the finalized design of a spacecraft \cite{ComparingDragCoeffs_GSIs, DSMCworkaroundDevised, dsmcFoam}.

In contrast to DSMC, analytical approaches such as closed-form equations take advantage of the physics of flight in VLEO to avoid particulate modeling. In VLEO, the atmosphere is rarefied, and the gas molecules have a large mean free path. Therefore, the number of inter-molecular collisions is small, and the atmospheric drag is dictated primarily by the interaction between the molecules and the surfaces of a spacecraft. The aforementioned GSIMs are mathematical descriptions of this effect. These models include equations that can be solved to calculate a body's drag coefficient, $C_d$. However, due to the complex mathematics involved, they quickly become prohibitively difficult for anything beyond basic geometric shapes such as spheres, flat plates, and cones. \cite{MoeMoe2005}.

The panel method involves breaking down complex geometries in such a way that the simple formulae of the GSIMs can be applied to obtain the total body drag and lift coefficients. They do not require the drastic simplification of complex shapes, as closed-form equations do. Thus, it can provide a much more accurate estimation of $C_d$ than closed-form equations for complex geometries. When compared to DSMC, the execution time for such programs is faster, and the computational load lower, due to the absence of simulated particles. Thus, they are more suited to deliver aerodynamic insight at the mission design stage. With an increasing number of satellites operating in VLEO, this will undoubtedly prove invaluable in the near future. Analysis of the geometries of recent significant missions to VLEO such as CHAMP \cite{CHAMP_firstpaper}, GRACE \cite{GRACE_firstpaper}, GOCE \cite{GOCE_firstpaper}, and SLATS \cite{SLATS} reveals a trend towards simple spacecraft shapes. 
Further aerodynamic considerations would require a thorough investigation of a wide array of design options, for which DSMC is limited in suitability due to its long runtime. Closed-form equations are also unsuitable due to their inability to capture the fine details of satellite bodies. Thus, ADBSat is better suited to this application than either of the other two methods available. 

While faster than DSMC, the accuracy of panel methods in VLEO has been hitherto largely unknown. ADBSat suffers from known limitations with regard to modeling phenomena such as multiple particle reflections between different components of the spacecraft body. Through validation, we have investigated a range of cases, both simplistic and realistic, and report here our findings with regards to scenarios for which it is particularly suited for analysis. A discussion is also presented of cases which display a significant inaccuracy. Comparison to DSMC constitutes our main method of validation, implemented through the open-source software suite OpenFOAM \cite{OFcitation} as dsmcFoam \cite{dsmcFoamPlus,dsmcFoam}. For simple geometries, cases are also compared to closed-form GSIM equations. Finally, we report the performance of ADBSat as compared to published literature results for real spacecraft. 

The comparison is complicated by the fact that ADBSat does not give an error margin on its outputs. Some uncertainty is expected due to factors such as the decomposition of the body into flat plates and the use of an atmospheric model rather than on-orbit data. These effects are difficult to quantify on a case-by-case basis. Therefore, we use the combination of ADBSat, dsmcFoam, and available literature to propose a fixed error margin, calculated as a percentage of $C_d$, that reflects the accuracy of ADBSat when compared to other methods of drag analysis. This captures some of the effects and prevents false confidence in what erroneously seems like an absolute result.

\section{Verification and validation methodology}
\label{S:2}

\subsection{Case equivalency between dsmcFoam and ADBSat}

We regarded dsmcFoam as the benchmark against which the results of ADBSat were tested. It is well-validated and known to be reliable for both transitional and rarefied gas flows \citep{dsmcFoamPlus, dsmcFoam4, dsmcFoam2}. It is also frequently maintained, with freely available documentation.

The DSMC algorithm relies on splitting the computational domain into many cells of side length $\Delta x$ and evolving particle motions by one time-step $\Delta t$ at a time. A key assumption is that each simulation particle, represented by a position vector \textbf{r} and velocity vector \textbf{V}, can represent many real particles \citep{dsmcFoam4}. The motion of particles is treated as being decoupled from collisions, such that the collisionless Boltzmann equation can be solved for each time-step. Once the motion has been propagated, the gas-surface and inter-molecular interactions are implemented.

The computational mesh is of utmost importance in DSMC. We incorporated CAD satellite geometries into the mesh using the \textit{blockMesh} and \textit{snappyHexMesh} utilities of the OpenFOAM package. Simulation parameters were chosen to satisfy a number of criteria:

\begin{enumerate}
    \item \textbf{Cell traverse time:} DSMC particles must not cross an entire simulation cell during one time step. Violation of this criterion could lead to artificial viscosity \citep{DSMCcriteria}.
    
    \begin{equation}
        v' \Delta t < \Delta x
        \label{eq:crit1}
    \end{equation}

   Where $v'$  is calculated using \cref{eq:mostProbSp}.
   
    \begin{equation}
    v' = \sqrt{\frac{2kT}{\bar{m}}}
    \label{eq:mostProbSp}
    \end{equation}

\item \textbf{Cell size:} $\Delta x$ must be on the order of, or smaller than $\lambda$. If this criterion is not fulfilled, there are many collisions per cell, resulting in the system approaching the continuum limit \citep{dsmcFoam,DSMCcriteria}.

    \begin{equation}
        \Delta x \leq \lambda
        \label{eq:crit2}
    \end{equation}

 $\lambda$ can be computed using \cref{eq:meanFreePath}.
    
    \begin{equation}
        \lambda = \frac{1}{n \pi d^2}
        \label{eq:meanFreePath}
    \end{equation}

\item \textbf{Collision timescale:} Similarly, $\Delta t$ must be smaller $\tau$, to maintain an appropriate number of collisions per cell.
    
    \begin{equation}
        \Delta t \leq \tau
        \label{eq:crit3}
    \end{equation}
    
    This is relatable to the previous criterion through \cref{eq:collTimescale}.
    
    \begin{equation}
        \lambda = \tau v'
        \label{eq:collTimescale}
    \end{equation}

\item \textbf{Simulation particle density, $\rho_n$:} The number of real molecules represented by a single simulation particle must be purposefully chosen. The program loses accuracy if $\rho_n$ is too low, due to an inflated number of collisions. Effectively, the result is an artificially lowered $K_n$ \citep{bird2007dsmc}. As computational time scales with $\rho_n^2$, values which are too high needlessly increase simulation time \cite{bird1994}. A suitable range is $7 \lesssim \rho_n \lesssim 20$.

\end{enumerate}

DsmcFoam applies the Maxwell GSIM, through the use of a Maxwellian thermal velocity distribution for the particles \cite{dsmcFoamPlus}. This allows the user to specify the fraction of the molecules that are diffusely re-emitted from the surface, with the remainder assumed to be specularly reflected. Reflection in VLEO with current typical spacecraft materials has been shown to be effectively diffuse \cite{MoeMoe2005}. Given a high degree of accommodation, which is observed in VLEO \citep{MoeMoe2005}, this GSIM produces results very close to those of a more complex model which more accurately reproduces molecular dynamics \cite{MaxwellVsCLL, MaxwellVsCLL2}. Therefore, the Maxwell model is appropriate for DSMC applications. However, this GSIM is not appropriate for panel methods, because of the fundamentally unrealistic way in which it describes the physics of non-equilibrium scattering events (those which differ from the average) \cite{GSIforDSMC}. These scattering events are integrated across the surface to obtain the mathematical expressions for $C_d$ utilized by ADBSat, which magnify this inaccuracy.
 
The GSIMs currently available in ADBSat are the Newton, Sentman \citep{sentman}, Schaaf and Chambre \citep{SchaafCham}, Cook \citep{Cook}, Cercignani-Lampis-Lord \cite{CLLmodel, CLL_Lord} and Storch \citep{storchHyp} models. The Newton model is also fundamentally inaccurate, and only included for estimation purposes. The Cook and Storch models would not yield the most general comparison to DSMC, as both are only applicable to the more limited case of hyperthermal flow. Of the remaining models, the CLL and Schaaf and Chambre models are dissimilar to the Maxwell model in their treatment of momentum accommodation, with both requiring two accommodation coefficients to the Maxwell model's one. This important distinction precludes the possibility of direct comparison. However, there are similarities between the Sentman and Maxwell models, with both using a Maxwellian velocity distribution function and assuming a diffuse re-emission profile. However, the Sentman equations include more physically realistic conditions through consideration of the random thermal motion of the molecules \cite{sabrina, sentman}. Thus, we selected the Sentman model to compare with dsmcFoam.

This model requires the accommodation coefficient, $\alpha$, to be specified. For the current available satellite materials, surfaces in VLEO are known to be contaminated with adsorbed oxygen. As a result, energy accommodation is assumed to be complete ($\alpha$ = 1) at altitudes up to \SI{200}{\kilo\metre}. This value decreases at higher altitudes, as surfaces become less contaminated \cite{MoeMoe1998, MoeMoe2005, AccommodationCoeffPardini}. For these cases, we calculated $\alpha$ by employing the model described by \citet{AccommodationCoeffModel}. The assumptions in this model align well with our simulations.It is most accurate at altitudes under \SI{500}{\kilo\metre}, making it particularly suitable for use in VLEO. While ADBSat has an input parameter to specify $\alpha$, dsmcFoam cannot account for partial accommodation. We used the work-around devised by \citet{DSMCworkaroundDevised} and implemented by \citet{DSMCworkaroundImplemented} to simulate incomplete accommodation in dsmcFoam. To paraphrase, we set the wall temperature of the satellite to be the temperature of the velocity distribution that the molecules would have, were they partly accommodated to the surface.

The accommodation coefficients we calculated are detailed in \cref{tab:accom} and broadly agree with those reported in other sources for similar altitudes \cite{MoeMoe1995, MoeMoe1998, MoeMoe2005, AccommodationCoeffPardini, AccommodationCoeffModel}. The satellite wall temperature at each orbital altitude considered is also presented in \cref{tab:accom}.

\begin{table}[hbt!]
\centering
\caption{Calculated parameters for the orbital altitudes, using the model described by \citet{AccommodationCoeffModel}.}
\label{tab:accom}
\begin{tabular}{c c c}
\hline
Altitude, km & $\alpha$ & $T_{k,r}$, K\\
\hline
100 & 1 & 300\\
200 & 1 & 300\\
300 & 0.97 & 1546.2 \\
400 & 0.81 & 7346.8 \\ 
\hline
\end{tabular}
\end{table}

In order to calculate these accommodation coefficients, we needed to determine the corresponding flight conditions. Verification and validation of ADBSat was intended to be as general as possible, hence, we chose intermediate solar activity conditions for the test cases, based on the $F_{10.7}$ solar index \cite{F10.7_definitions}. We chose a reference date and location of 19 January 2015 at midnight, latitude and longitude (0,0). Explicitly, 81-day average $F_{10.7} = 138.1$ and daily $F_{10.7} = 121.7$. $A_p$ magnetic indices were in the range \SIrange{2.9}{9.0}{}. We took the orbital velocity of the body as a function of altitude to be the free-stream velocity, $V_\infty$. 

ADBSat requires these inputs from the user in order to translate them to atmospheric parameters, by applying the US Naval Research Laboratory's mass spectrometer and incoherent scatter radar model (NRLMSISE-00) \cite{NRLMSISE00}. While other atmospheric models exist, such as the Jacchia-Bowman model (JB2008) \cite{JB2006, JB2008} and NASA's drag temperature model (DTM) \cite{DTM2000, DTM2013}, this model was chosen for three reasons: 

\begin{enumerate}
\item It has been compared to the other models \cite{NRLMSISE00, JB_NRLMSISE_DTM_comparison, DTM2009} and to real satellite data \cite{NRLMSISE_rocket_data, EelcoThesis}. It was shown to be at least as good as the other models at reproducing realistic atmospheric conditions \cite{EelcoThesis}.
\item There is an ongoing effort to maintain and improve the model. Corrections include employing new experimental data to correct the outputs \cite{NRLMSISE_champ_correction, NRLMSISE_2, cheng2019density, 2019NNimprovement}.
\item It is available as a MATLAB package, and is therefore easy to integrate into ADBSat.
\end{enumerate}

The data from the NRLMSISE-00 model can also be found online\footnote{Available at: \url{https://ccmc.gsfc.nasa.gov/modelweb/models/nrlmsise00.php}, accessed 16/07/2021} alongside its source code\footnote{Available at: \url{https://www.brodo.de/space/nrlmsise/}, accessed 16/07/2021}. By utilizing both sources, we maintained consistency between the manually specified DSMC atmospheric conditions, downloaded through the online tool, and the ones accessed by ADBSat through MATLAB. Thus, a direct comparison between the two methods of drag analysis was facilitated.

\subsection{Analysis of dsmcFoam outputs}

To calculate the aerodynamic forces, dsmcFoam integrates the pressure and skin-friction forces over a specified boundary, which in our case is the CAD geometry. This calculation is performed at each output time-step. The user must process the resulting set of $C_d$ values. 

\begin{figure}[hbt!]
\centering\includegraphics[width=0.95\linewidth]{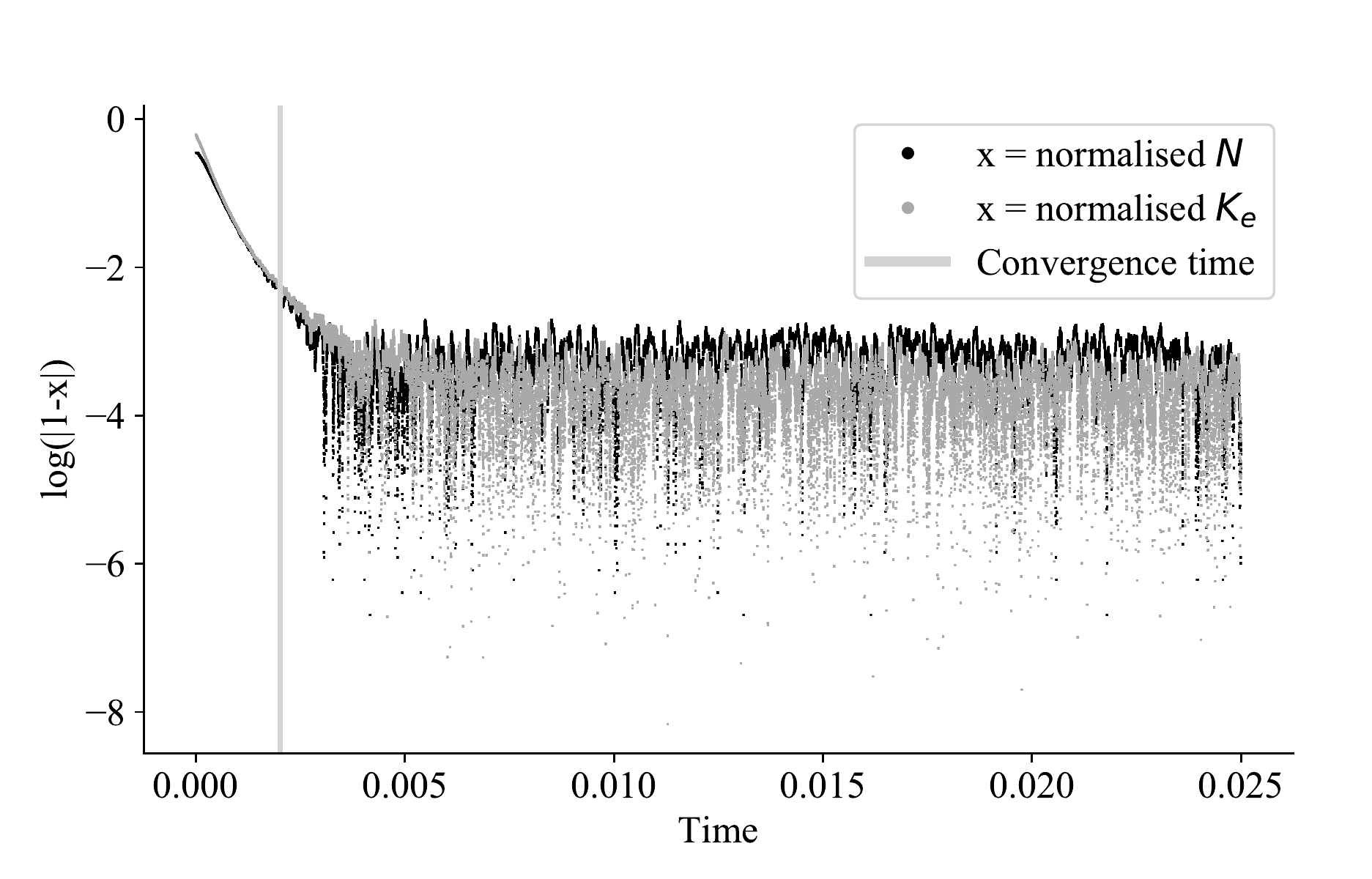}
\caption{Analysis of dsmcFoam convergence for a sphere at \SI{200}{\kilo\metre} altitude.}
\label{fig:conv_analy}
\end{figure}

Steady-state convergence is necessary for accuracy \citep{dsmcFoam} i.e., the total number of simulation particles $N$ and the total linear kinetic energy $K_e$ must plateau. We ran the simulations until plots of $N$ and $K_e$ showed no significant change over the preceding 10,000 time-steps. To determine the convergence time $t_c$, we then used the final value of $N$ as a reference and compared every other value in the set to this. We determined the $t_c$ as the time-step at which the value reached within $\pm$0.5\% of the reference. A similar analysis was conducted for $K_e$. We used the later time-step of the two as the point of convergence. All values of $C_d$ output before this point were discarded. \Cref{fig:conv_analy} shows a graph of values of $N$ and $K_e$ for the simulation of a sphere at \SI{200}{\kilo\metre}, scaled to the reference. The convergence time of \SI{0.00202}{\second} is shown using a vertical gray line, which is equivalent to 4040 time-steps. The details of convergence for the entire sample of test cases is shown in \cref{tab:conv}, with "rejected" indicating the percentage of values after $t_c$ which did not lie within $\pm$0.5\% of the final value.

\begin{table}[hbt!]
\centering
\caption{Details of the convergence of the entire test sample.}
\begin{tabular}{c | c | c | c}
\hline
 & \textbf{$t_c$} & \textbf{$\Delta t_c$} & \textbf{rejected (\%)}\\
\hline
Minimum & 0.00019 & 380 & 0 \\
Maximum & 0.00508 & 10160 & 1.69\\
Mean & 0.00208 & 4168 & 0.1054  \\
Median & 0.00193 & 3860 & 0.0299 \\
\end{tabular}
\label{tab:conv}
\end{table}

\begin{figure}[!t]
\centering\includegraphics[width=0.8\linewidth]{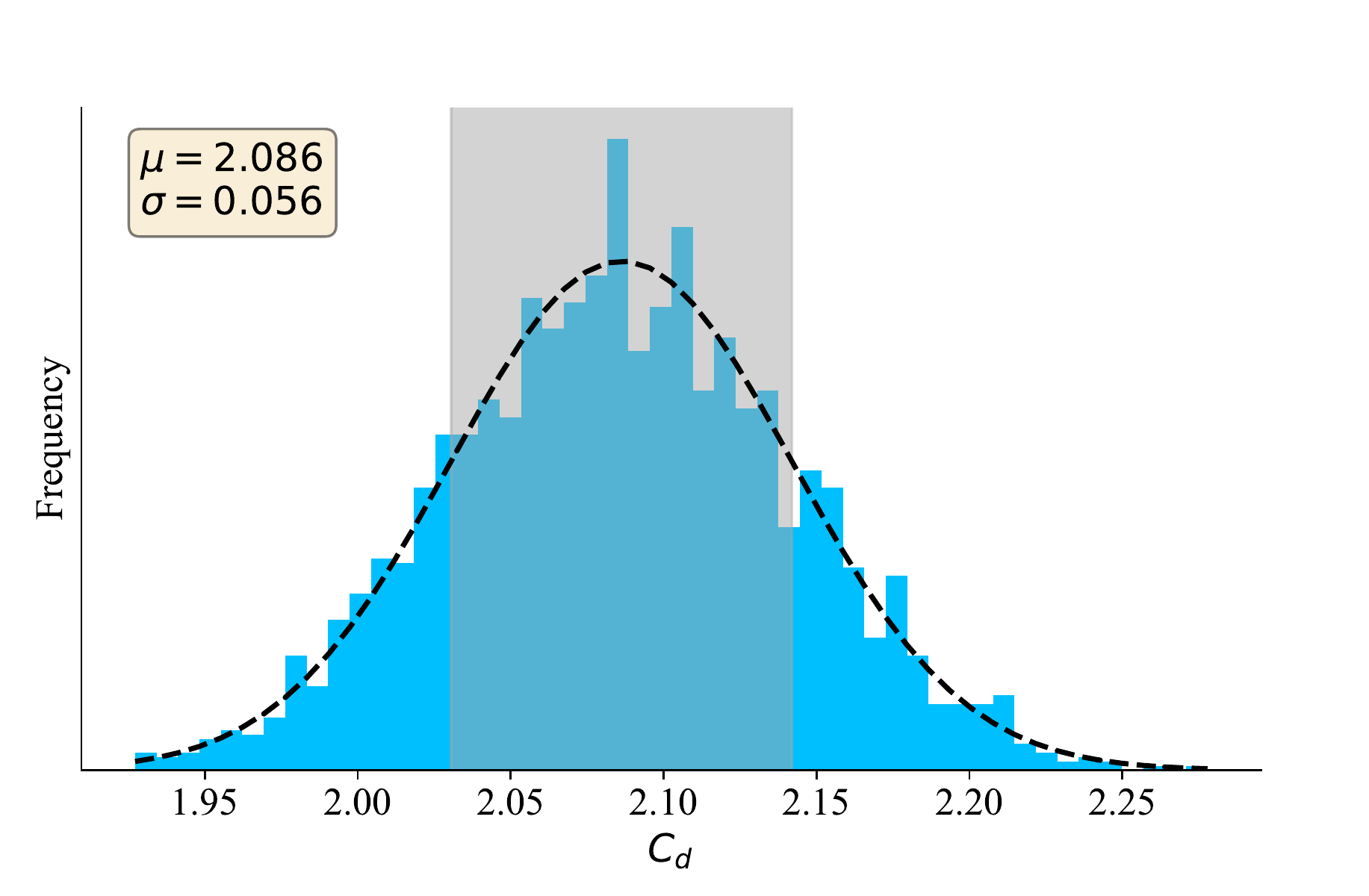}
\caption{A histogram of $C_d$ values for a sphere at \SI{200}{\kilo\metre}, alongside the corresponding Gaussian function. The $1\sigma$ range is highlighted in gray.}
\label{fig:Gaussian}
\end{figure}

As DSMC is a stochastic method \cite{bird1994}, each measure of $C_d$ is instantaneous, directly related to the individual particle positions and velocities at the time step at which it is calculated. Thus, the values exhibit scatter, due to the fluctuation of instantaneous force with time. This random scatter forms an approximately Gaussian distribution, as shown in \cref{fig:Gaussian}, where the Gaussian function based on $\mu$ and $\sigma$ in the top left corner is shown as a dashed black line. We used $\mu$ as the final value of $C_d$. We interpreted $\sigma$, shown in \cref{fig:Gaussian} in light gray, as the error on the mean. In later figures, where $\sigma$ is large enough to be shown graphically, it is represented by error bars.

\subsection{Description of Test Cases}

\begin{figure}[t]
\centering\includegraphics[width=\linewidth]{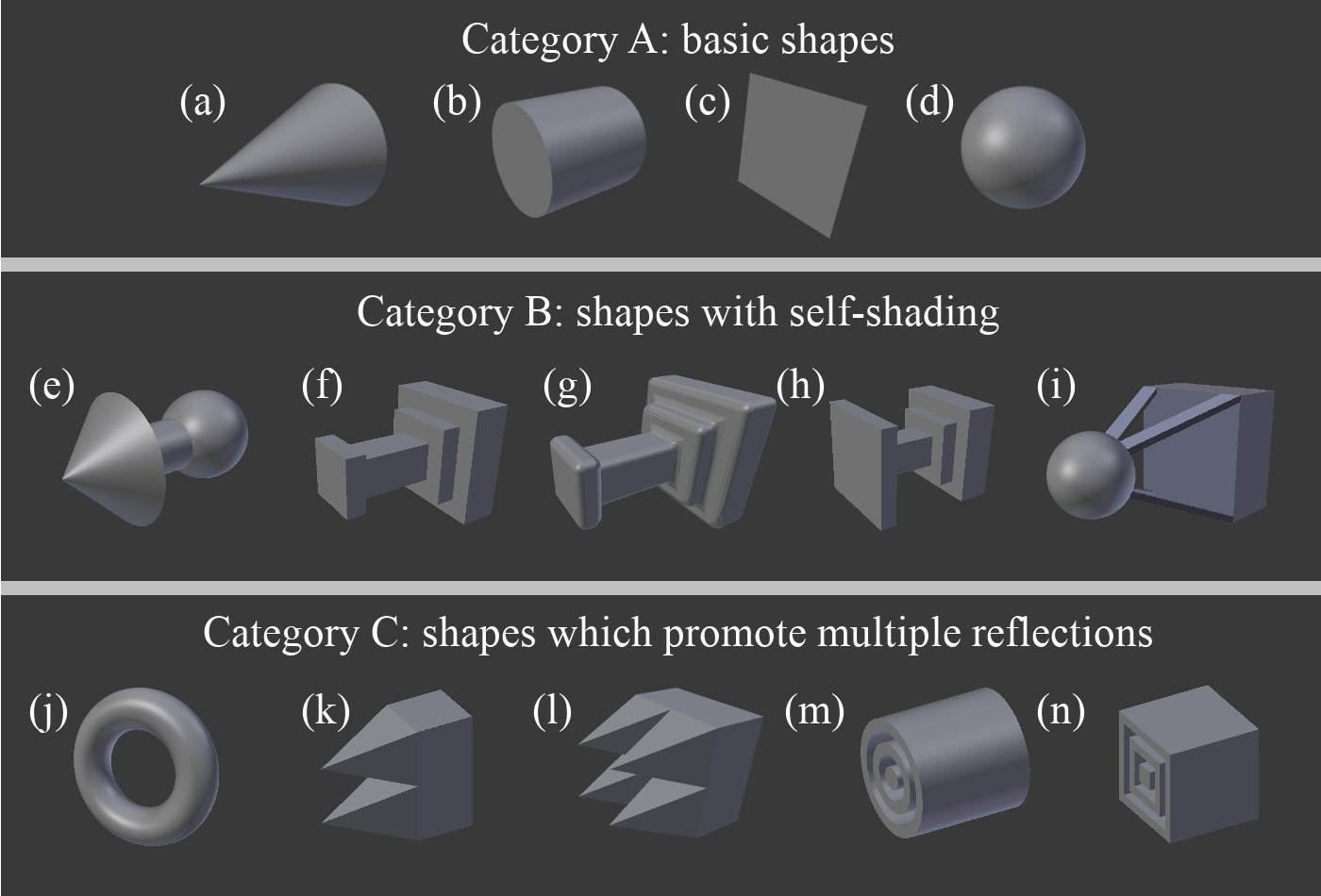}
\caption{Selected shapes used for testing.}
\label{fig:shapes_collage}
\end{figure}

We devised a number of initial shapes to test specific aspects of ADBSat, as shown in \cref{fig:shapes_collage}. They range from \SIrange{0.2}{0.9}{\metre} in length, with height and width being of the same order of magnitude. We categorized them as follows:

\begin{enumerate}
    \item \textbf{Category A}: basic shapes, (a) to (d) in \cref{fig:shapes_collage}. The Sentman GSIM provides closed-form solutions not only for a flat plate with one side exposed to the flow, but for four additional basic geometries \cite{sentman}. We chose these shapes so that results from ADBSat could also be compared to these solutions, as a secondary check of basic functionality. 
    
    \item \textbf{Category B}: shapes with self-shading, (e) to (i) in \cref{fig:shapes_collage}. These shapes test ADBSat's shading algorithm by having some panels shielded from the flow by upwind features of the body.
    
    \item \textbf{Category C}: shapes which promote multiple particle reflections, (j) to (n) on \cref{fig:shapes_collage}. These shapes employ details such as angled panels or concavities on the forward-facing surfaces to promote reflection of the particles between faces.
\end{enumerate}

We examined shapes in category A at orbital altitudes from \SIrange{100}{400}{\kilo\metre}, in intervals of \SI{100}{\kilo\metre}. All other shapes were examined at an altitude of \SI{200}{\kilo\metre} only. We set AOA and AOS to zero for all shapes. Additionally, we also examined category B shapes a second time at a random selection of small AOA and AOS, ranging from 2.8 to 12.1 degrees, as a secondary test of the shading algorithm. Furthermore, these shapes were also discretised into smaller panels than the default CAD geometry, by manually selecting a target face on each shape and subdividing it. Our aim was to ascertain the influence of panel size on the output value of $C_d$.

\begin{figure}[hbt!]
\centering\includegraphics[width=0.85\linewidth]{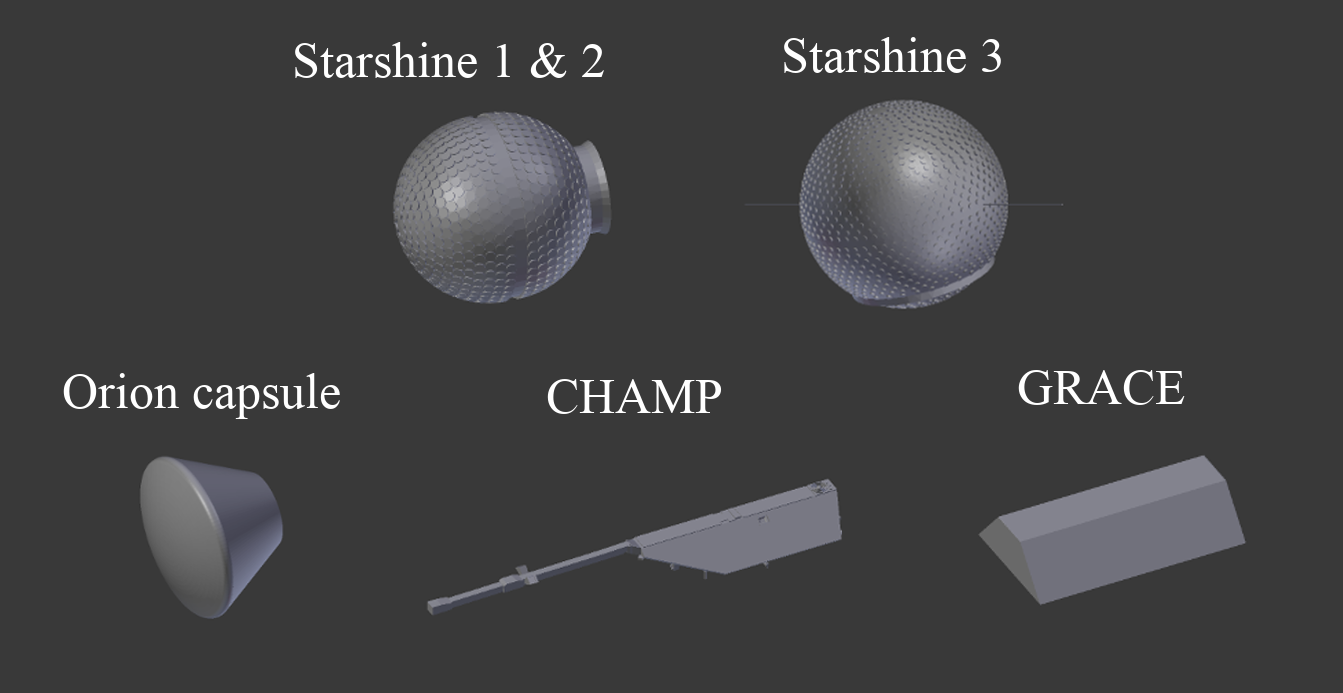}
\caption{Models of the real satellites that were examined.}
\label{fig:sats_collage}
\end{figure}

Finally, we identified literature sources that offered drag coefficient data for real satellite missions. The methods employed varied across sources, comprising of analytical equations, DSMC, and free-molecular code. We reproduced as best as possible the simulation conditions reported, in order to directly compare our results to the data. The five objects examined, each at a range of conditions, can be seen in \cref{fig:sats_collage}. We built the CAD models of the Orion capsule \cite{orion} and simplified GRACE satellite \cite{responseSurfacesMehta} from technical drawings. The CHAMP geometry is the high-fidelity model of the spacecraft produced by \citet{HighFidelityModels}. Both Starshine satellite geometries are those used by \citet{DSMCworkaroundDevised}, and were obtained from the author via personal communication.

In actuality, we considered many other sources on the subject of aerodynamic analysis of real satellites, but many did not provide the data required to reproduce their results. Key simulation parameters were omitted which made comparison impossible. As a guide, \cref{tab:key_sim_params} outlines parameters which are necessary, and those which are advantageous, for reproducing published results.

\begin{table}[hbt!]
\centering
\caption{Simulation parameters which are necessary or advantageous for reproducibility of results.}
\begin{tabular}{c | c }
\hline
\textbf{Necessary} & \textbf{Advantageous}\\
\hline
Choice of GSIM & Atmospheric particle density\\
GSIM parameters, such as $\alpha$ & Atmospheric temperature\\
Choice of atmospheric model & Mean molecular mass\\
Altitude & Diagrams of the objects\\
Solar activity levels & Free-stream velocity\\
Reference cross-sectional area & Knudsen number\\
\end{tabular}
\label{tab:key_sim_params}
\end{table}

\section{Results}

\subsection{Category A (basic shapes)}

\begin{figure}[hbt!]
     \centering
     \begin{subfigure}[c]{0.45\textwidth}
         \centering 
         \includegraphics[width=\textwidth,height=5cm]{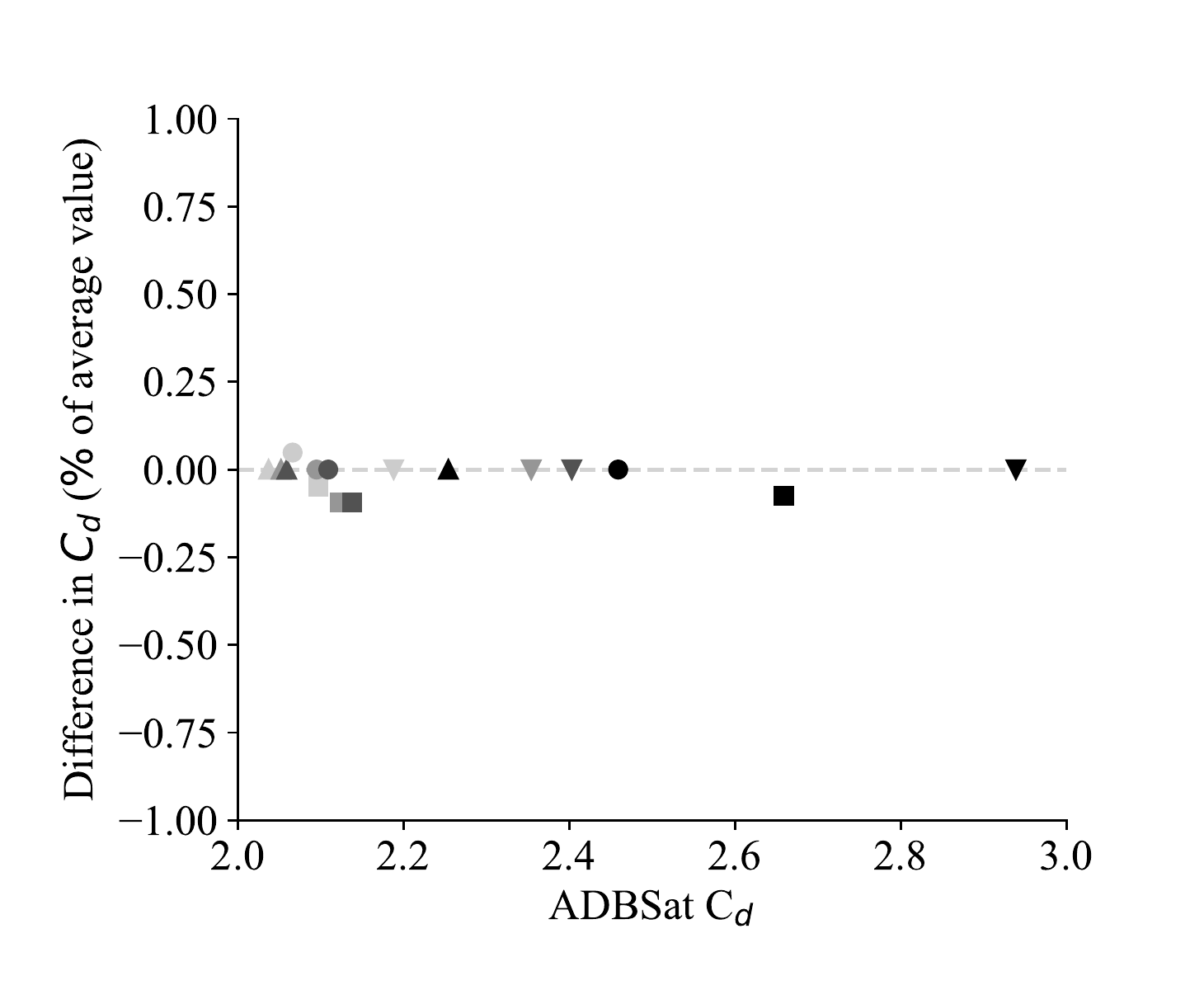}
         \caption{ADBSat vs Sentman equations}
         \label{fig:basic_ADBsat_Sentman}
     \end{subfigure}
     \hfill
     \begin{subfigure}[c]{0.54\textwidth}
         \centering
         \includegraphics[width=\textwidth,height=4.75cm]{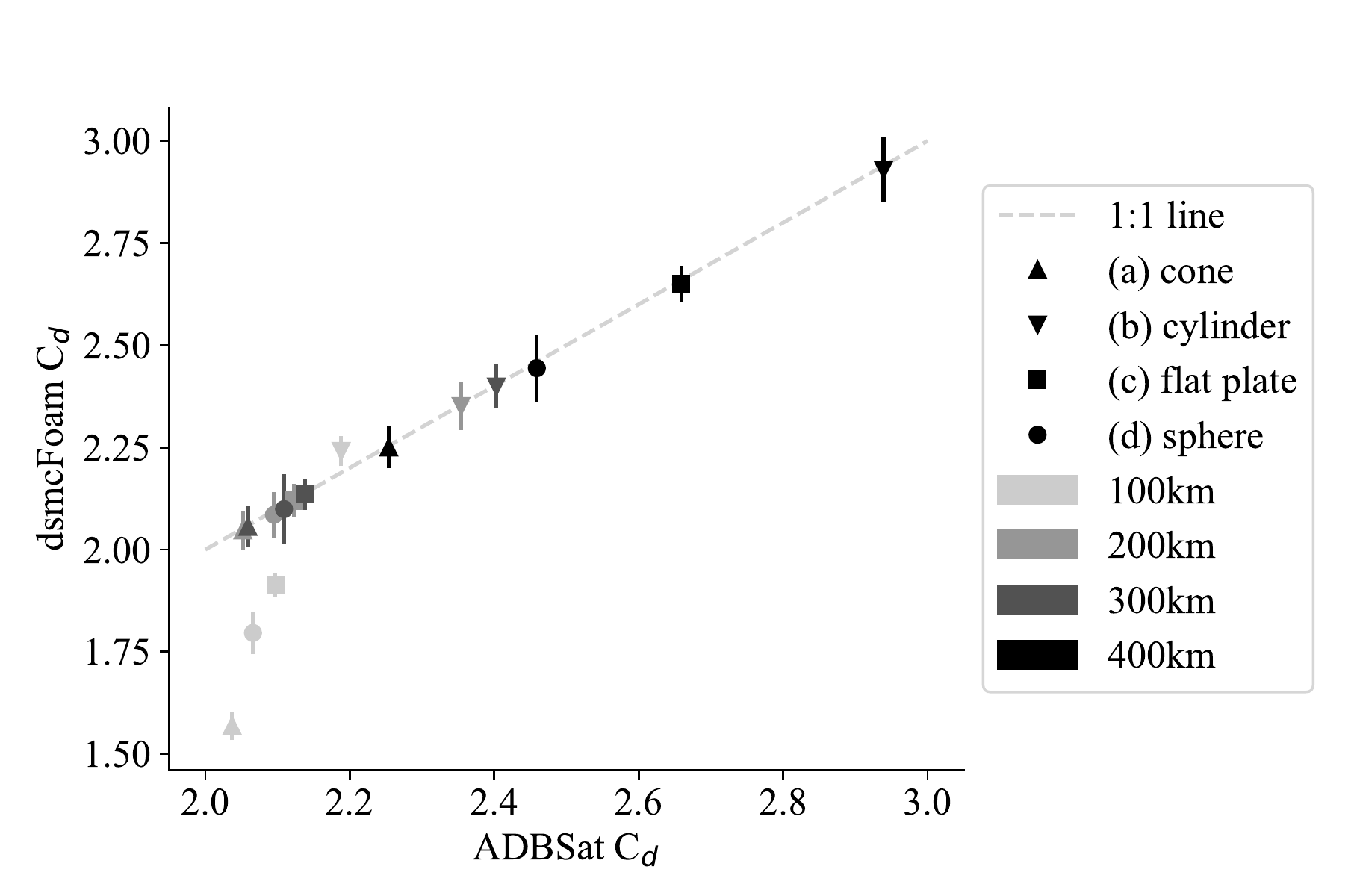}
         \caption{ADBSat vs dsmcFoam}
         \label{fig:basic_ADBSat_dsmcFoam}
     \end{subfigure}
        \caption{Comparison between ADBSat, the Sentman closed-form equations, and dsmcFoam. Note the small scale of the y-axis in \cref{fig:basic_ADBsat_Sentman}.}
        \label{fig:basic_shapes}
\end{figure}

We first compared the results from ADBSat with those from the closed-form Sentman GSIM equations. It was expected that they would agree closely, as they are essentially two different methods of applying the same theory. Indeed, no case shows a difference in $C_d$ higher than 0.1\% between the two methods, well within the expected error limits of ADBSat. The percentage difference across the samples can be seen in \cref{fig:basic_ADBsat_Sentman}, which shows a good agreement between the two methods. The slightly higher errors on the flat plate cases are due to the Sentman model assuming a zero thickness plate, which ADBSat cannot model. A small non-zero thickness was modeled instead.

Secondly, we verified results from ADBSat against those obtained from dsmcFoam. Runtime analysis revealed that that, for basic shapes, ADBSat is approximately five orders of magnitude faster than dsmcFoam: where dsmcFoam needs about \SI[parse-numbers=false]{10^5}{\second} to run a single simulation, ADBSat completes the same analysis in less than \SI{10}{\second}. Running multiple simulations, such as those at varying AOA and AOS that are required to obtain the aerodynamic database of a satellite, will compound this difference further. Thus, there is a clear time advantage to using ADBSat over dsmcFoam. \Cref{fig:basic_ADBSat_dsmcFoam} shows a graph of the values of $C_d$ obtained from these two methods. Results are consistent for all shapes, at all orbital altitudes except \SI{100}{\kilo\metre}. 

For the atmospheric parameters chosen, at an altitude of \SI{100}{\kilo\metre}, $n_0 \sim 10^{17}$ and $K_n \sim 0.5$. Thus, the flow has become transitional, and the equations applied by ADBSat do not reflect the physics involved. However, they perform well at higher altitudes, where strict FMF conditions exist. This is consistent with existing literature which addresses the comparison of closed-form equations to DSMC \cite{ComparingDragCoeffs_GSIs}. As a result of the natural fluctuation of atmospheric conditions, attempting to define a lower limit of accuracy for ADBSat in terms of height would be misguided. In conclusion, we have shown that the panel method applied in ADBSat is consistent with dsmcFoam for basic convex geometries across all altitudes where $K_n \ge 10$.

\subsection{Category B (shading algorithm)}

\begin{figure}[b!]
     \centering
     \begin{subfigure}[c]{0.38\textwidth}
         \centering
         \includegraphics[width=\textwidth,height=5cm]{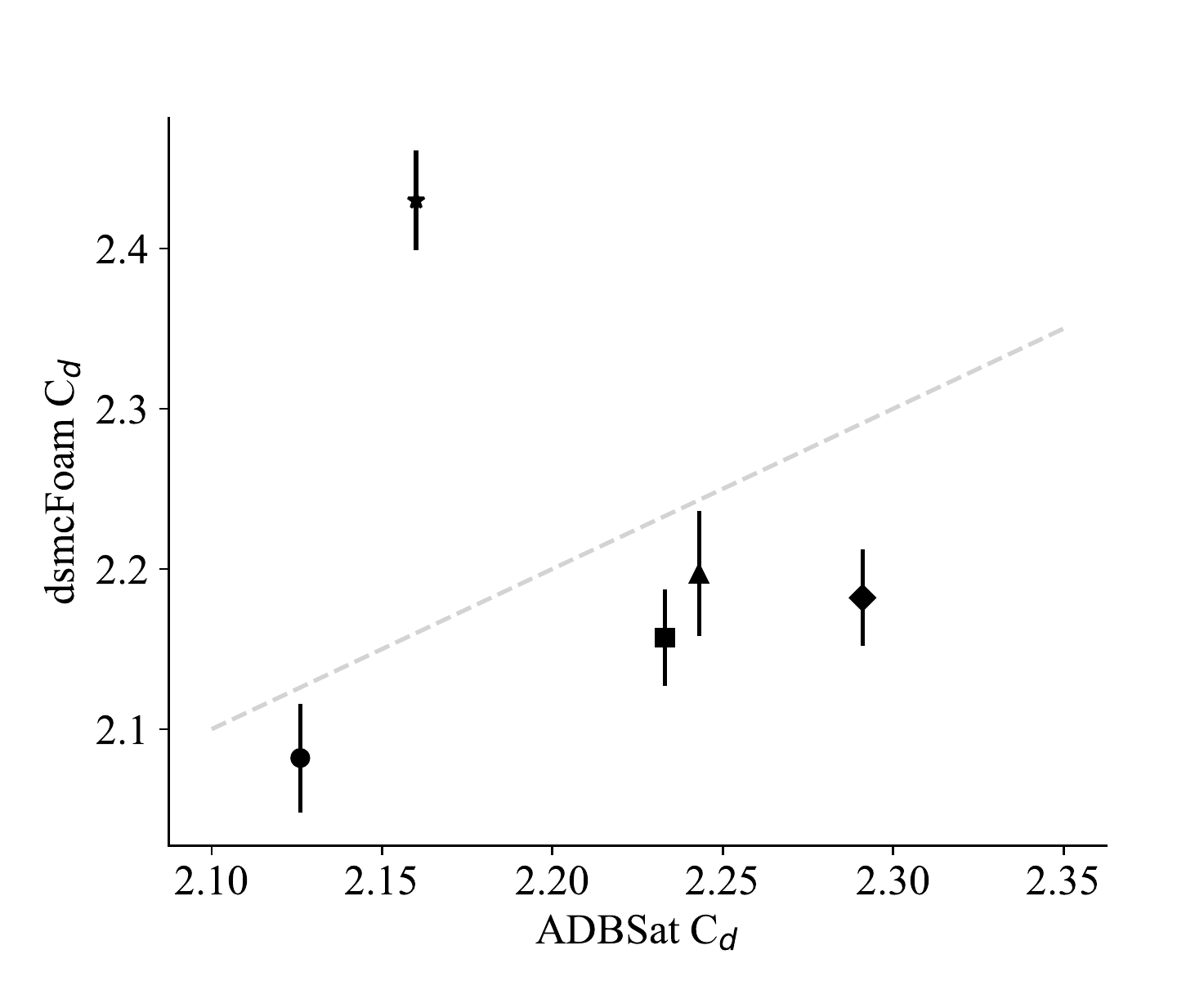}
         \caption{shading, no subdivision}
         \label{fig:shading_nosubd_headon}
     \end{subfigure}
     \hfill
     \begin{subfigure}[c]{0.61\textwidth}
         \centering
         \includegraphics[width=\textwidth,height=5cm]{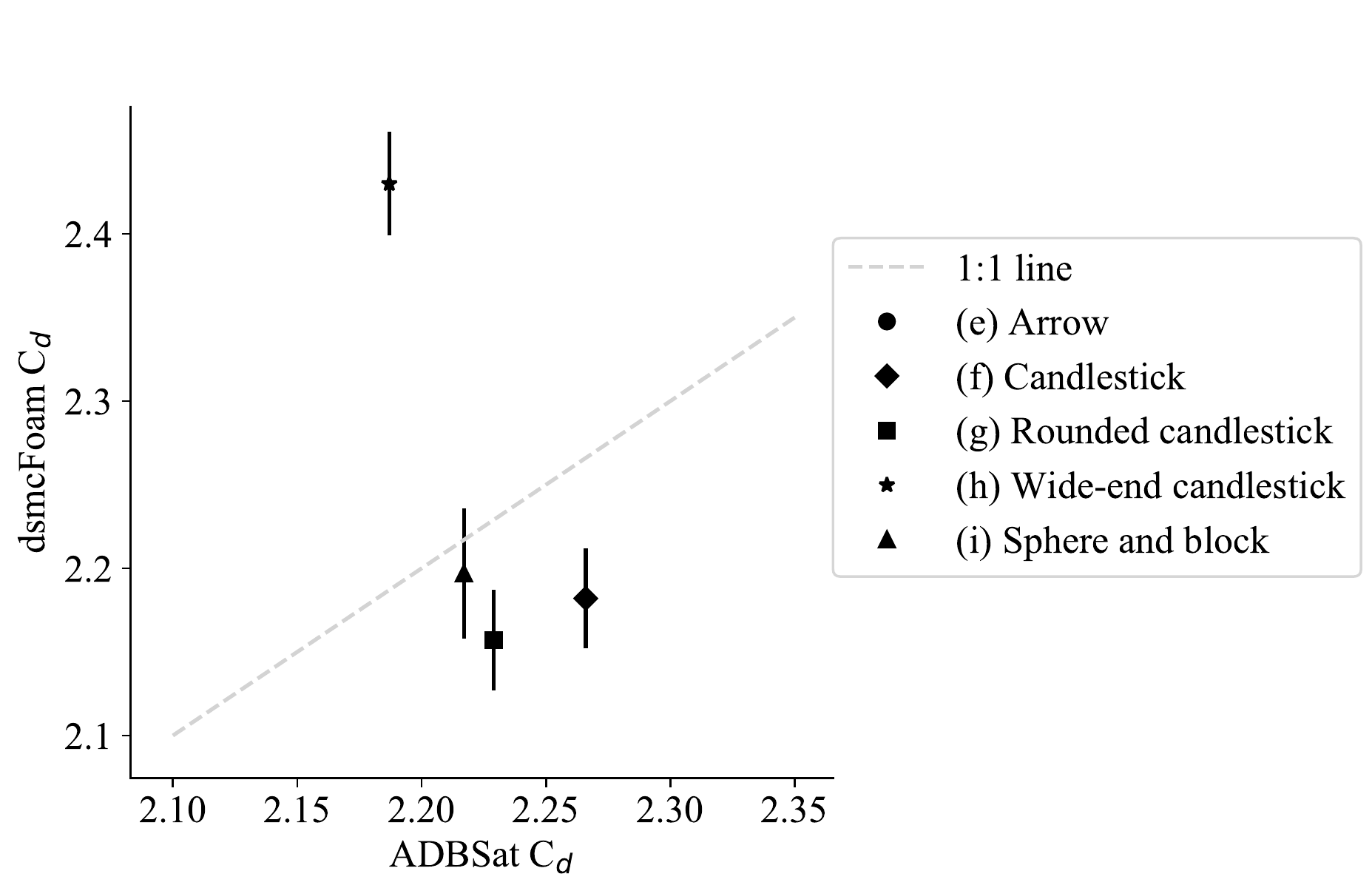}
         \caption{shading with subdivision}
         \label{fig:shading_subd_headon}
     \end{subfigure}
        \caption{ADBSat vs dsmcFoam results for category B, for head-on flow. }
        \label{fig:shading_headon}
\end{figure}

Having established that ADBSat performs well for simple shapes, we examined Category B shapes to ascertain the accuracy of the shading algorithm. The results of this analysis for head-on flow can be seen in \cref{fig:shading_headon}. 

Initially, we represented the shapes using the fewest possible panels, in order to keep computational time to a minimum. However, as seen in \cref{fig:shading_nosubd_headon}, ADBSat yielded no results that agree with dsmcFoam. While two shapes show results within 2$\sigma$, with others showing as much as an 8.7$\sigma$ difference, it is clear that the two methods are not consistent.

\begin{figure}[t!]
\centering
     \begin{subfigure}{0.4\textwidth}
         \centering
         \includegraphics[width=\textwidth]{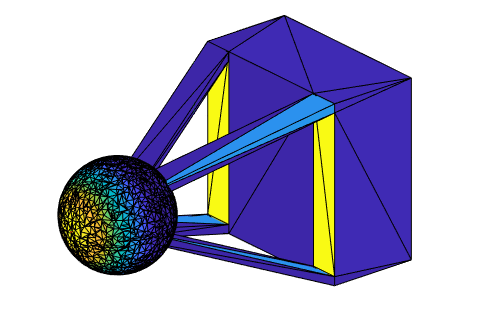}
         \caption{Pre-subdivision}
         \label{fig:dragAlong_nosubd}
     \end{subfigure}
    \
     \begin{subfigure}{0.4\textwidth}
         \centering
         \includegraphics[width=\textwidth]{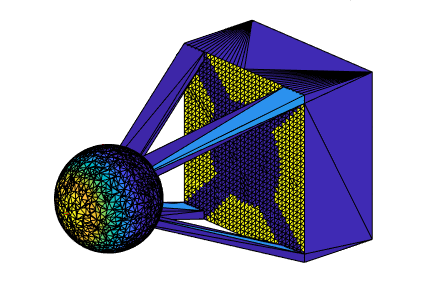}
         \caption{Post-subdivision}
         \label{fig:dragAlong_subd}
     \end{subfigure}
        \caption{Shape (i) before and after manual subdivision, with the flow head-on to the spherical feature. Blue regions contribute much less to the total $C_d$ than yellow.}
        \label{fig:manual_discretisation}
\end{figure}

We also manually subdivided some of the large flat sides of shapes (f)-(i) into smaller panels during the CAD design process, in the hope that this would capture more realistic shading effects. This was unnecessary for shape (e), as the curved nature of the body was discretised into small panels by default. We saw some improvement in all the results, with no significant difference in runtime. However, only one was within 1$\sigma$ of the value from dsmcFoam: shape (i). It exhibited unrealistic results for the shading prior to subdivision, which improved after, as seen in \cref{fig:manual_discretisation}. For other shapes, the difference before and after subdivision is less apparent. Therefore, while this effect does not fully account for the observed discrepancy, it can have a significant influence on the analysis of complex shapes.

\begin{figure}[hbt!]
    \centering
    \includegraphics[width=0.58\linewidth]{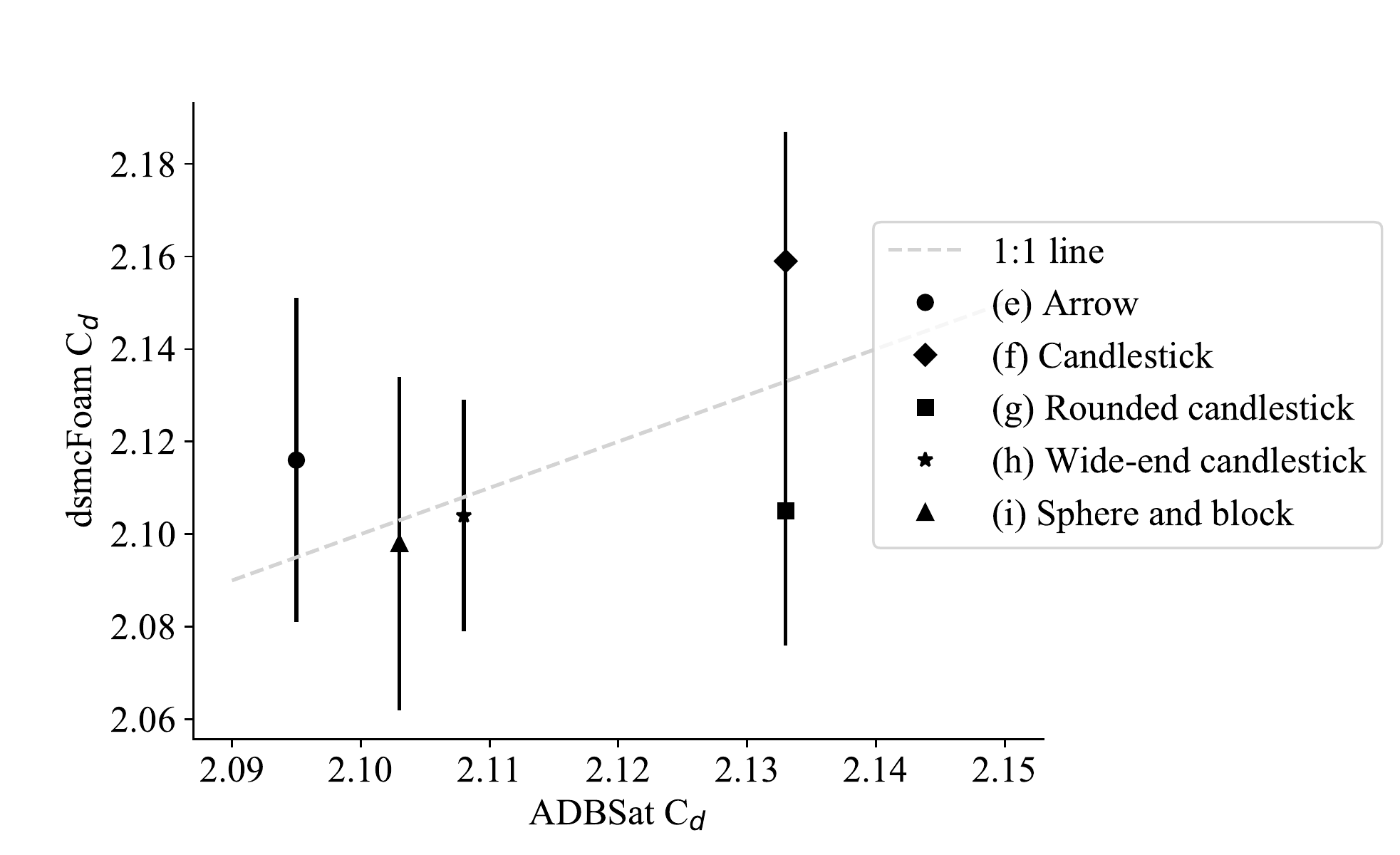}
    \caption{Analysis of the shading algorithm performance, for category B shapes at non-zero AOA and AOS.}
    \label{fig:shading_angled}
\end{figure}

\begin{figure}[b!]
     \centering
     \begin{subfigure}[h]{0.38\textwidth}
         \centering
         \includegraphics[width=\textwidth]{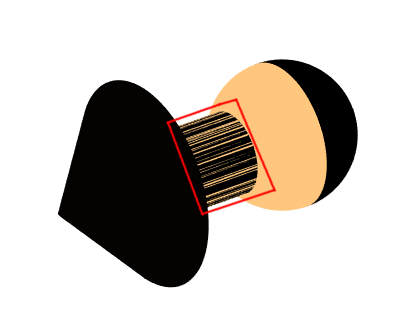}
         \caption{Shading for shape (e)}
         \label{fig:arrow_wrongshading}
     \end{subfigure}
     \
     \begin{subfigure}[h]{0.38\textwidth}
         \centering
         \includegraphics[width=\textwidth]{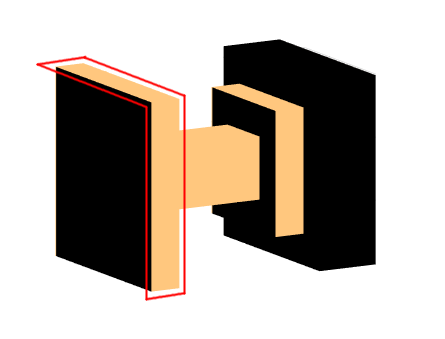}
         \caption{Shading for shape (h)}
         \label{fig:bigpawn_wrongshading}
     \end{subfigure}
        \caption{Erroneously categorized panels with regard to shading, highlighted in red.}
        \label{fig:shading_effect}
\end{figure}

In contrast, the same set of shapes analyzed at an angle yields more consistent results between the two methods, as seen in \cref{fig:shading_angled}. All five geometries now agree within 1$\sigma$ with dsmcFoam. As each shape was examined at a different combination of random AOA and AOS, it is clear that the previous issues are seen only for head-on flow, when many panels are at an angle of $\delta \sim$ \SI{90}{\degree} to the flow and to each other. The cause of this is the inherent computational error of MATLAB, caused by the representation of the geometry using floating point values. When panels are at an angle of \SI{90}{\degree} to each other and to the flow, the 2D projection employed by the shading algorithm will project the barycenter of the downwind shaded panel exactly on the edge of the upwind shading panel. Thus, the barycenter is neither inside nor outside the shading panel. However, the shading algorithm must still decide which side of the edge it falls on. The small computational error which MATLAB carries will influence such a precise calculation heavily, resulting in an almost random classification of shaded/non-shaded panels. A visualization of this effect is seen in \cref{fig:shading_effect}. Panels classified by the program as shaded are in orange, with those not shaded in black. The edges of each panel are not outlined, for clarity. The area containing incorrectly classified panels is outlined in red. \Cref{fig:arrow_wrongshading} shows shape (e), for which all panels of the central cylindrical section should be shaded. \Cref{fig:bigpawn_wrongshading} shows shape (h), for which the edge panels highlighted should not be shaded. As the shaded panels are effectively removed from the final summation, too few shaded panels lead to an overestimation of $C_d$, while too many shaded panels will lead to underestimation.

However, when the flow is at an angle of $\delta \ge$ \SI{0.1}{\degree}, the projection of the shaded panel's barycenter is no longer on the edge of the shading panel. Thus, MATLAB's small error is now negligible, and ADBSat yields much more accurate results. In summary, ADBSat can handle the majority of satellite flight scenarios with accurate shading analysis, except those where many large panels are at an angle of $\delta \approx$ \SI{90}{\degree} to the flow and to each other. As the CAD model of any geometry must be made independently of ADBSat, it should be apparent to the user whether or not shading analysis is required, and if yes, whether any AOA and AOS will pose an problem. For such cases, we recommend interpolating between multiple values of $C_d$ that span across the problem case.

\subsection{Category C (multiple reflections)}

\begin{figure}[hbt!]
\centering
\includegraphics[width=0.5\linewidth]{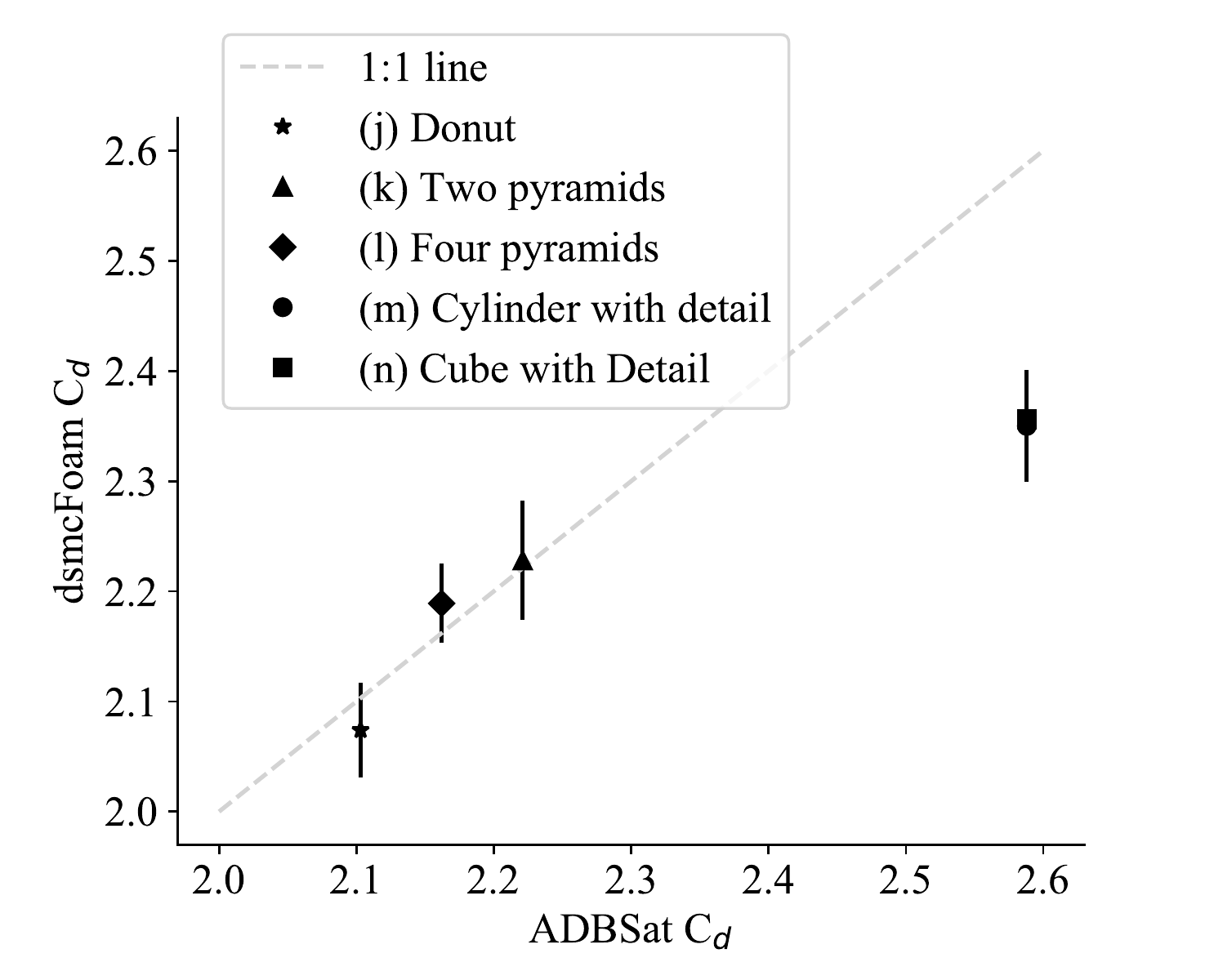}
\caption{ADBSat vs. dsmcFoam results for category C shapes.}
\label{fig:mult_reflections}
\end{figure}

The comparison of category C shapes can be seen in \cref{fig:mult_reflections}. The aim was primarily to ascertain the influence of multiple particle reflections. For the three simpler shapes, (j), (k) and (l), the results of ADBSat and dsmcFoam are in agreement. However, the two more detailed shapes, (m) and (n), show discrepancy between the two methods. 

Under the assumption of diffuse re-emission, multiple particle reflections should not affect the overall result considerably, due to the shape of the reflected distribution. Unlike specular reflection, it has an element of randomness in the particle velocities \cite{DragModelling}. We have determined that for simpler shapes, the effect is small enough that the results output by ADBSat are within the error limits of dsmcFoam.

However for shapes (m) and (n), the deep grooves of the forward face lead to particles being trapped in the indentations. This means that any panels inside the grooves contribute to $C_d$ in a fundamentally different manner than the single particle reflections which ADBSat assumes. ADBSat essentially treats shape (m) as a cylinder with extra surface area perpendicular to the flow, caused by the sides of the indentations. Thus, it shows a higher $C_d$ for this shape than for the plain cylinder. In contrast, dsmcFoam can capture the effect of the trapped particles and their reflections more accurately. It can take into account the effects of the localized increase in particle number - and thus, pressure and $K_n$ - inside the grooves, which ADBSat cannot. dsmcFoam therefore finds a value of $C_d$ for shape (m) closer to that of the unchanged cylinder. A similar analysis also applies for shape (n).

In summary, multiple reflections of particles between the surfaces of the body do not lead to a large inaccuracy, for relatively shallow features. However, ADBSat is unsuitable for satellite shapes that include deep features where particle trapping could occur, such as intakes.

\subsection{Starshine satellites}

\begin{figure}[h!]
     \centering
     \begin{subfigure}[c]{0.45\textwidth}
         \centering
         \includegraphics[width=\textwidth, height=5cm]{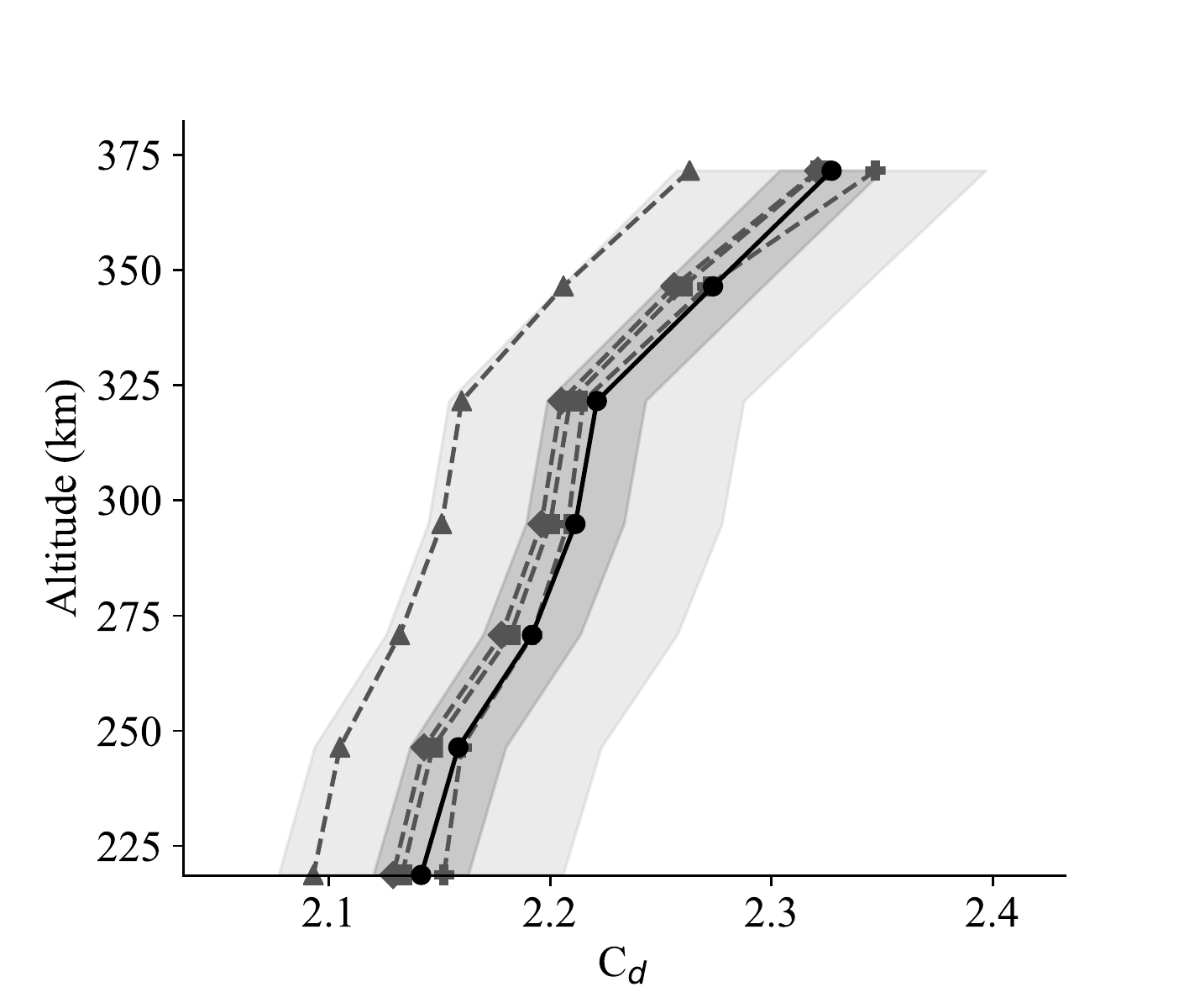}
         \caption{Starshine 1}
         \label{fig:starshine1}
     \end{subfigure}
\
     \begin{subfigure}[c]{0.45\textwidth}
         \centering
         \includegraphics[width=\textwidth, height=5cm]{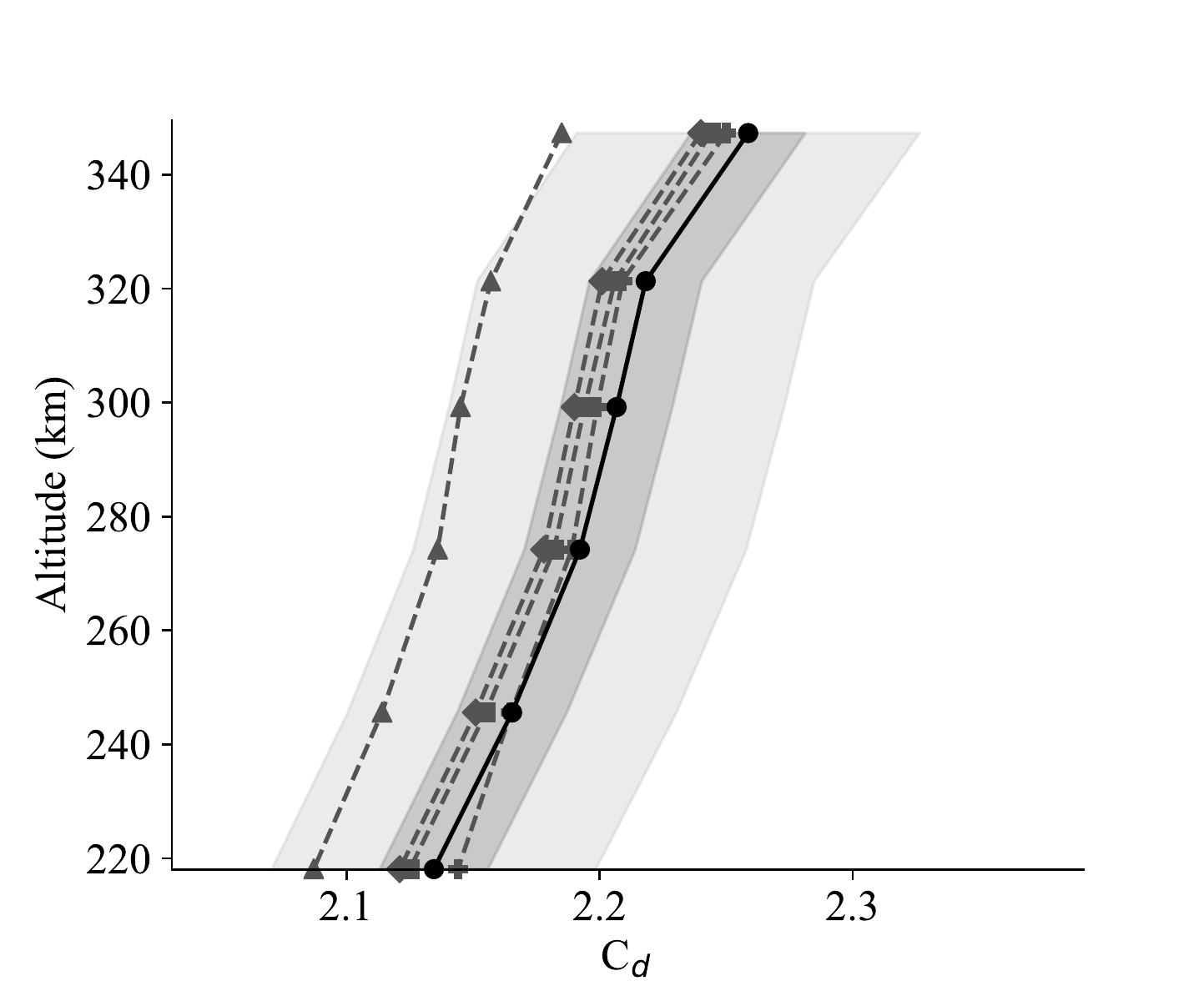}
         \caption{Starshine 2}
         \label{fig:starshine2}
     \end{subfigure}
     
     \begin{subfigure}[c]{0.45\textwidth}
         \centering \hspace{-0.7cm}
         \includegraphics[width=\textwidth, height=5cm]{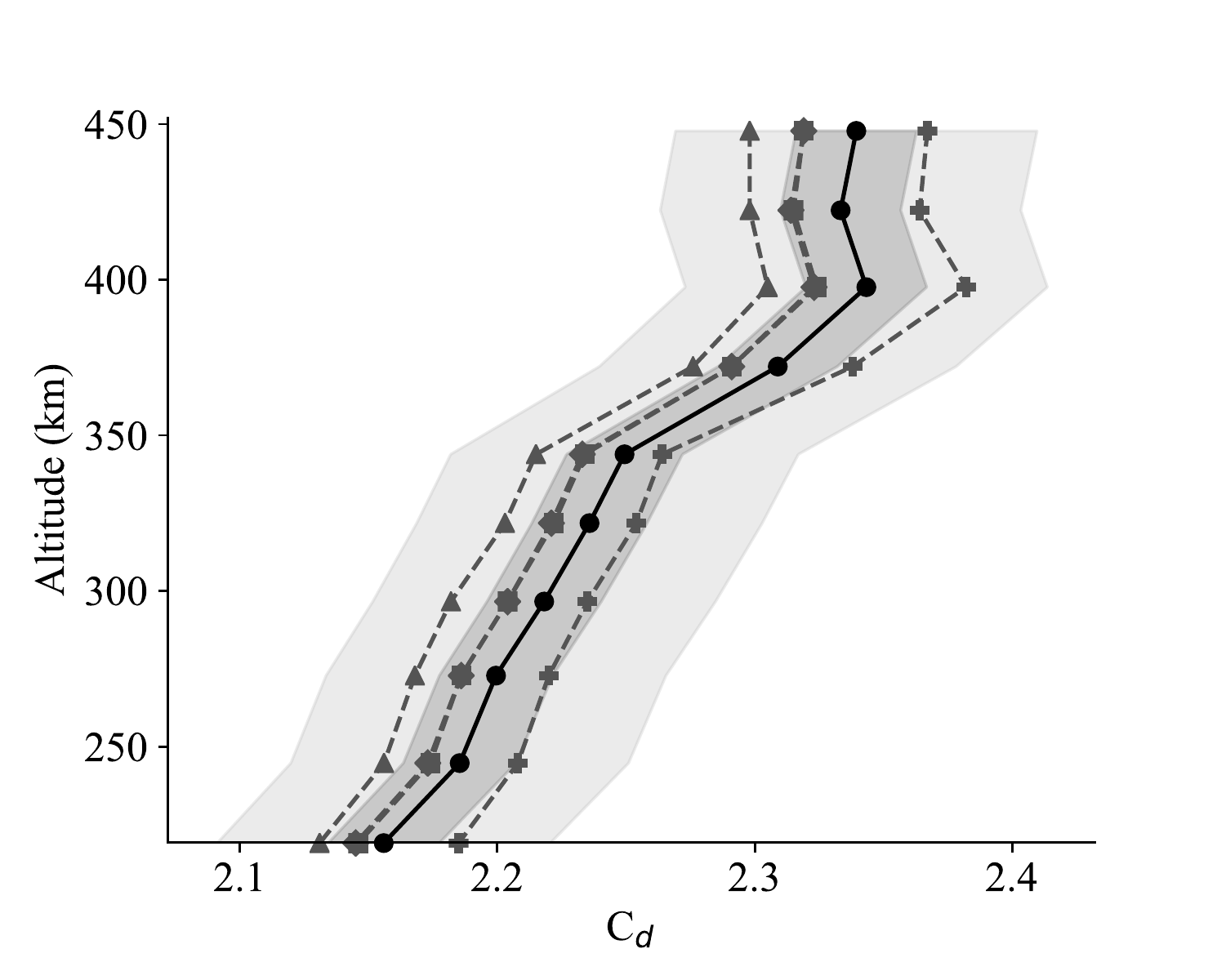}
         \caption{Starshine 3}
         \label{fig:starshine3}
     \end{subfigure}
\hspace{1cm}     
     \begin{subfigure}[c]{0.3\textwidth}
         \centering
         \vspace{-1cm} \hspace{-1cm}
         \includegraphics[width=\textwidth]{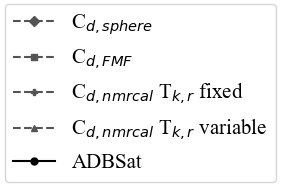}
     \end{subfigure}
     
        \caption{Comparison between $C_d$ output by ADBSat and that presented by \citet{DSMCworkaroundDevised}, for the three Starshine satellites across a range of altitudes. Error ranges of 1\% and 3\% on the ADBSat results are highlighted.}
        \label{fig:starshineAltitudes}
\end{figure}

\citet{DSMCworkaroundDevised} used the \textit{DS3V} DSMC code for their drag analysis of the Starshine satellites. As we were able to obtain the CAD geometry files they used, we are confident of the validity of this comparison. 

Their results for the variation of $C_d$ with altitude, for all three satellites, are shown in \cref{fig:starshineAltitudes} alongside our own. They employed multiple methods of calculating $C_d$:

\begin{figure}[h!]
\centering
\includegraphics[width=0.8\linewidth]{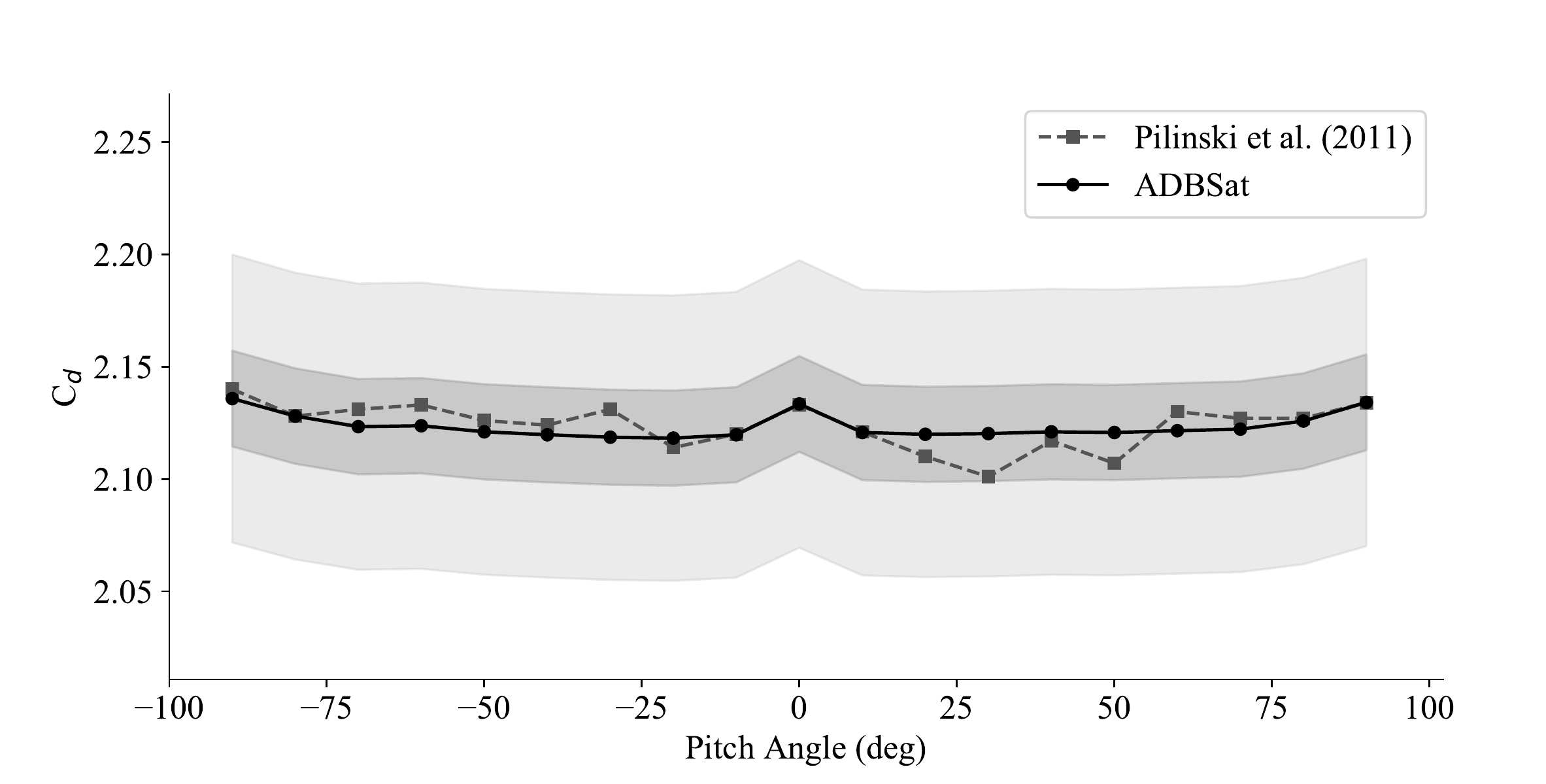}
\caption{$C_d$ at each pitch angle for Starshines 1 and 2, as calculated by \citet{DSMCworkaroundDevised} and ADBSat. Note the fine scale of the y-axis. The gray bands indicate error ranges of 1\% and 3\% on ADBSat.}
\label{fig:starshineAngle}
\end{figure}

\begin{figure}[h!]
     \centering
     \begin{subfigure}{0.45\textwidth}
         \centering
         \includegraphics[width=\textwidth]{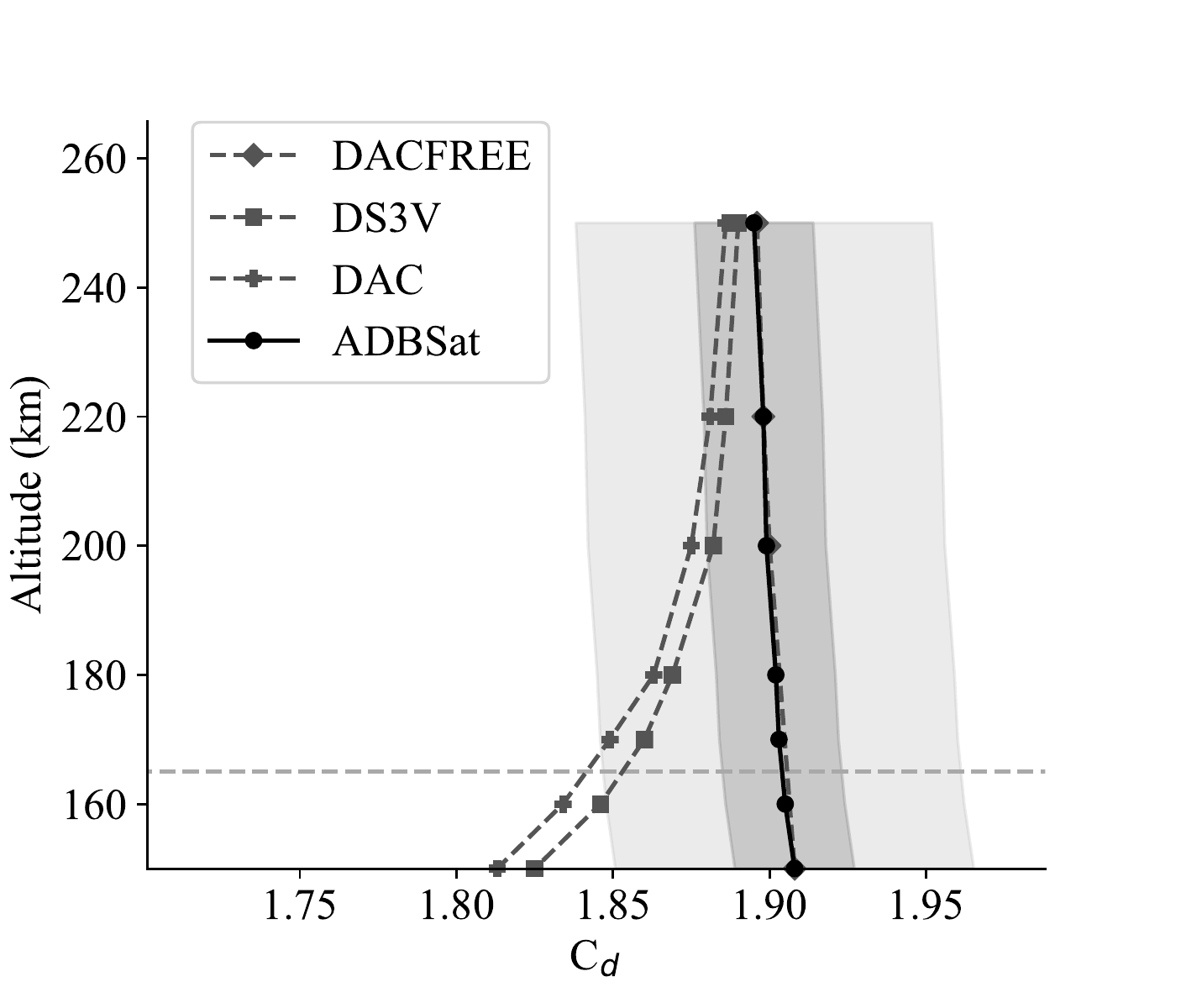}
         \caption{Comparison across altitudes}
         \label{fig:orion_alts}
     \end{subfigure}
\
     \begin{subfigure}{0.4\textwidth}
         \centering
         \includegraphics[width=\textwidth]{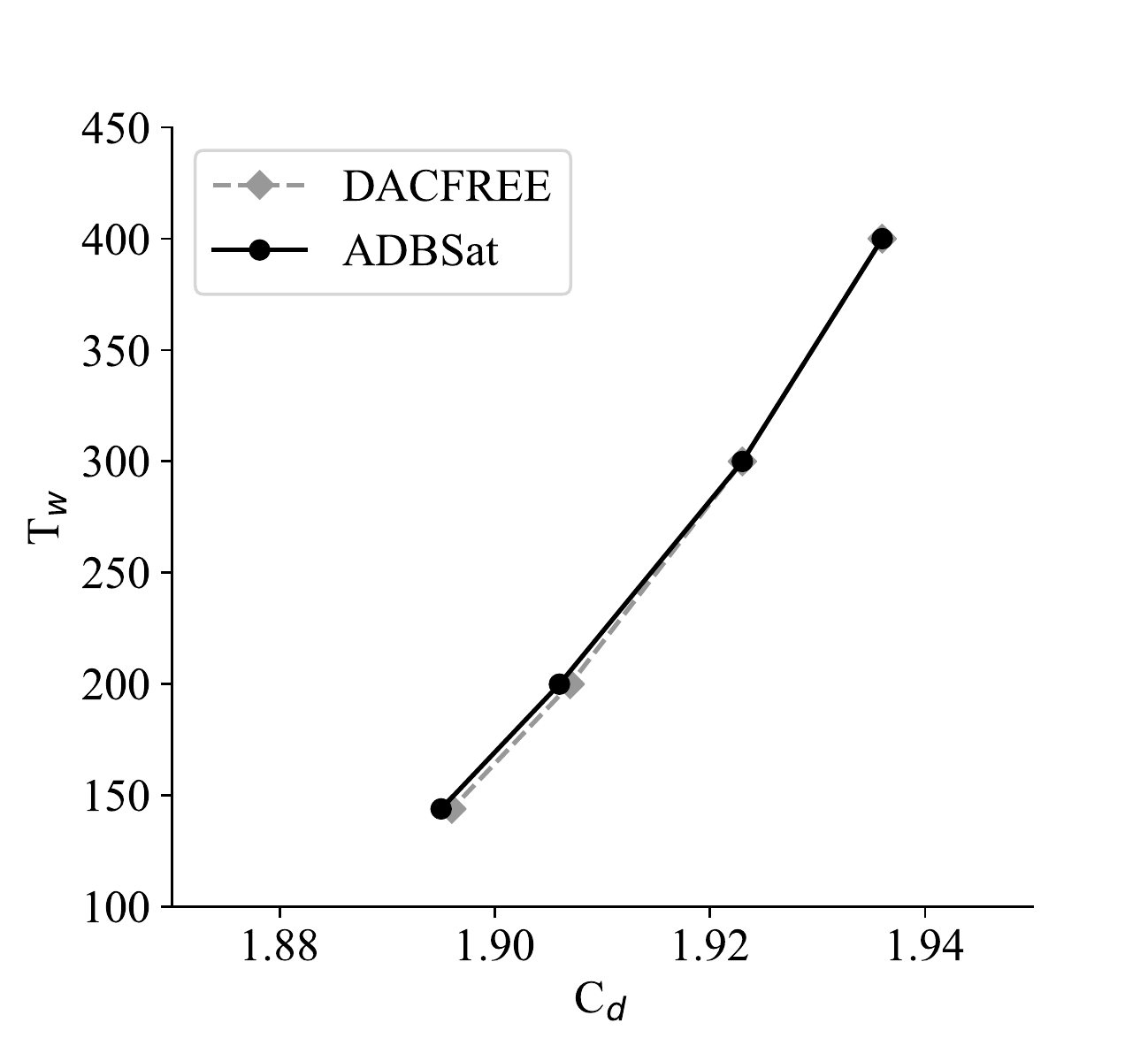}
         \caption{Comparison across $T_w$}
         \label{fig:orion_Tw}
     \end{subfigure}
     
        \caption{Comparison between ADBSat and \citet{orion} of drag analysis for the Orion capsule. The 1\% and 3\% error intervals on the ADBSat outputs are highlighted in \cref{fig:orion_alts}.}
        \label{fig:orion}
\end{figure}

\begin{enumerate}
    \item $C_{d,sphere}$: closed-form equations for the drag on a perfect sphere
    \item $C_{d,FMF}$: a panel method which computes the drag of each mesh element without considerations of shadowing or multiple reflection
    \item $C_{d,nmrcal}$ $T_{k,r}$ $fixed$: a test-particle Monte Carlo (TPMC) method with single-impact accommodation
    \item $C_{d,nmrcal}$ $T_{k,r}$ $variable$: the same TPMC with multiple reflections
\end{enumerate}

ADBSat agrees well with the reported values of $C_{d,sphere}$, $C_{d,FMF}$ and $C_{d,nmrcal}$ $T_{k,r}$ $fixed$. An error of 1\% (highlighted in dark gray) comfortably encompasses their data. Setting the error to 3\% (highlighted in light gray) also covers the more realistic case of multiple reflections.

Furthermore, we also verified our results against their aerodynamic analysis of Starshines 1/2 as a function of pitch angle, shown in \cref{fig:starshineAngle}. ADBSat shows a similar trend, with slightly less fluctuation, than the DSMC calculations. All values are within 1\%, indicating good agreement between the two methods of aerodynamic analysis.

\subsection{Orion capsule}

The detailed description of atmospheric and geometric parameters provided by \citet{orion} allowed us to accurately reproduce their simulations. The authors present results obtained with three different codes, two DSMC codes (\textit{DAC} and \textit{DS3V}) and a free-molecular code whose algorithm is not known (\textit{DACFREE}). Their values of $C_d$ at relevant altitudes are reproduced in \cref{fig:orion}, alongside results from ADBSat. The dark and light grey highlighted areas are the 1\% and 3\% intervals on ADBSat values, respectively.

ADBSat agrees closely with \textit{DACFREE} for all simulations, even those not in FMF. At approximately \SI{170}{\kilo\metre} and above, where FMF occurs according to the $K_n$ reported by \citet{orion}, it also agrees to within 1-3\% with both DSMC codes. A better agreement is noted at higher altitudes, where the flow is strictly FMF. To summarize, our results show a good agreement to the $C_d$ presented, where the assumption of FMF is valid.

\subsection{GRACE}

\begin{figure}[hbt]
\centering\includegraphics[width=0.5\linewidth]{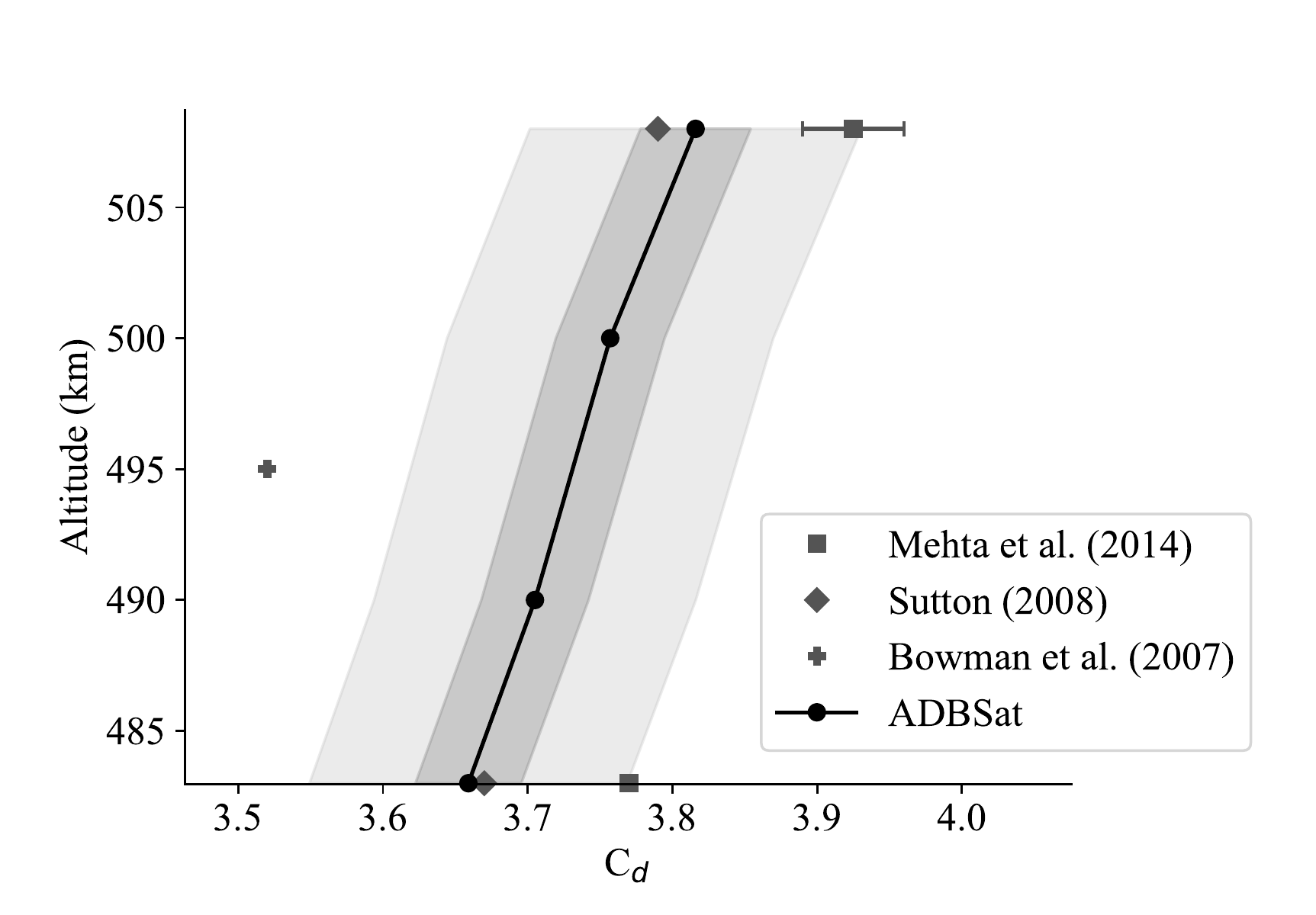}
\caption{ADBSat vs. literature sources for the drag coefficient of the GRACE satellite. 1\% and 3\% regions on ADBSat outputs are highlighted.}
\label{fig:GRACE}
\end{figure}

\citet{responseSurfacesMehta} report the variation in drag coefficient of GRACE over \SI{24}{\hour}, both from their own data and from previous sources. The satellite's orbit over this time will vary from \SIrange{483}{508}{\kilo\metre} continually. For comparison purposes, we assumed that the maximum and minimum drag coefficients of \citet{responseSurfacesMehta} and \citet{SuttonThesis} occur at the apogee and perigee respectively. The average ballistic coefficient found by \citet{bowmanGRACECHAMP} was assumed to be at the midpoint of the altitude range, \SI{495}{\kilo\metre}.

ADBSat calculated values of $C_d$ closest to those reported by \citet{SuttonThesis}, as shown in \cref{fig:GRACE}. The agreement is comfortably within 1\%. The values reported by \citet{responseSurfacesMehta} are within a wider margin of 3\% of our results. The average reported by \citet{bowmanGRACECHAMP} is lower than our value, most likely due to similar reasons as those reported in \citet{responseSurfacesMehta}, such as their use of a constant accommodation coefficient as opposed to to our altitude-dependent analysis.

\subsection{CHAMP}

\begin{figure}[hbt]
     \centering
     \begin{subfigure}{0.6\textwidth}
         \centering
         \includegraphics[width=\textwidth]{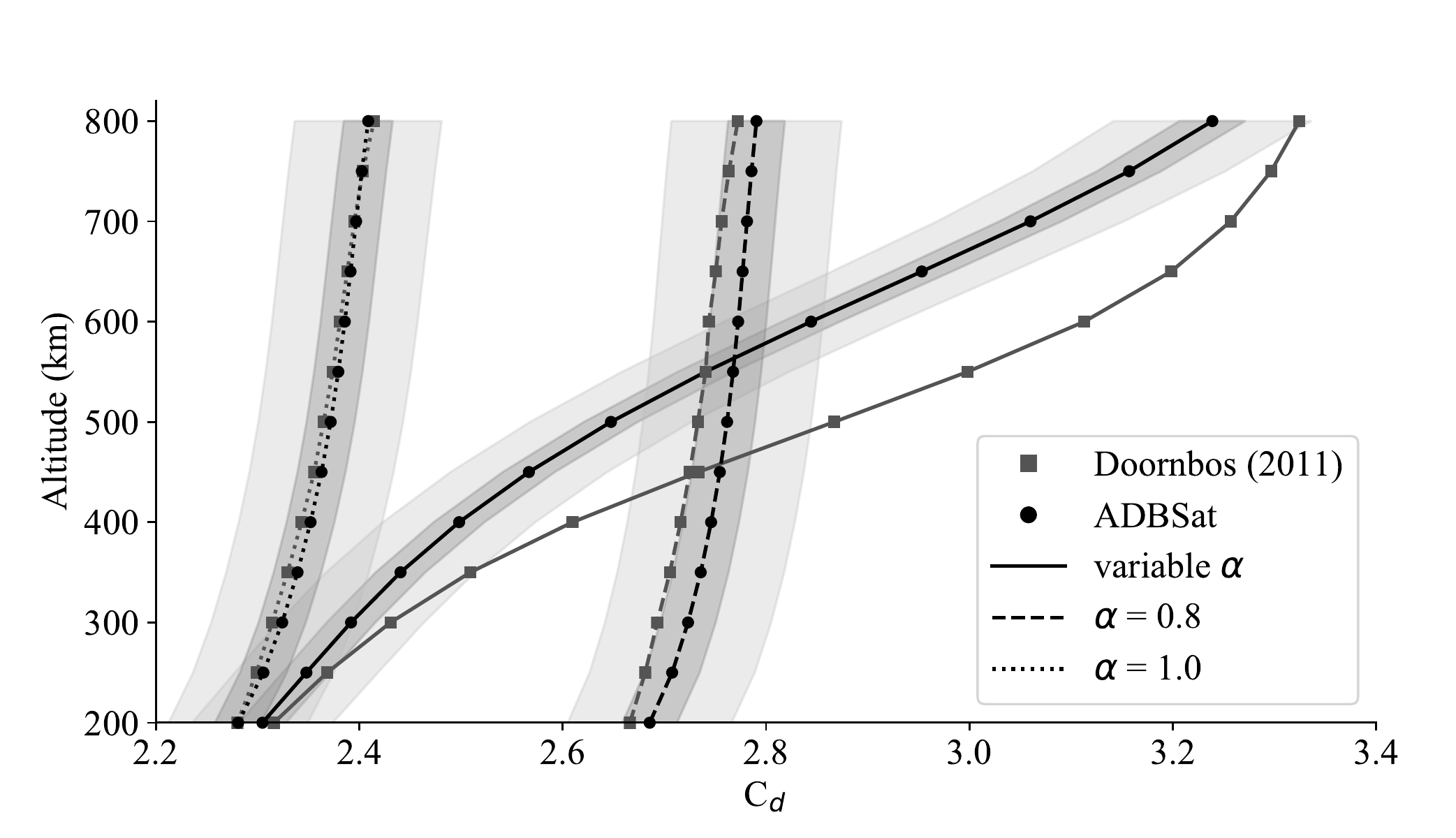}
         \caption{Solar maximum}
         \label{fig:CHAMP_solmax}
     \end{subfigure}
     
     \begin{subfigure}{0.6\textwidth} 

         \centering
         \includegraphics[width=\textwidth]{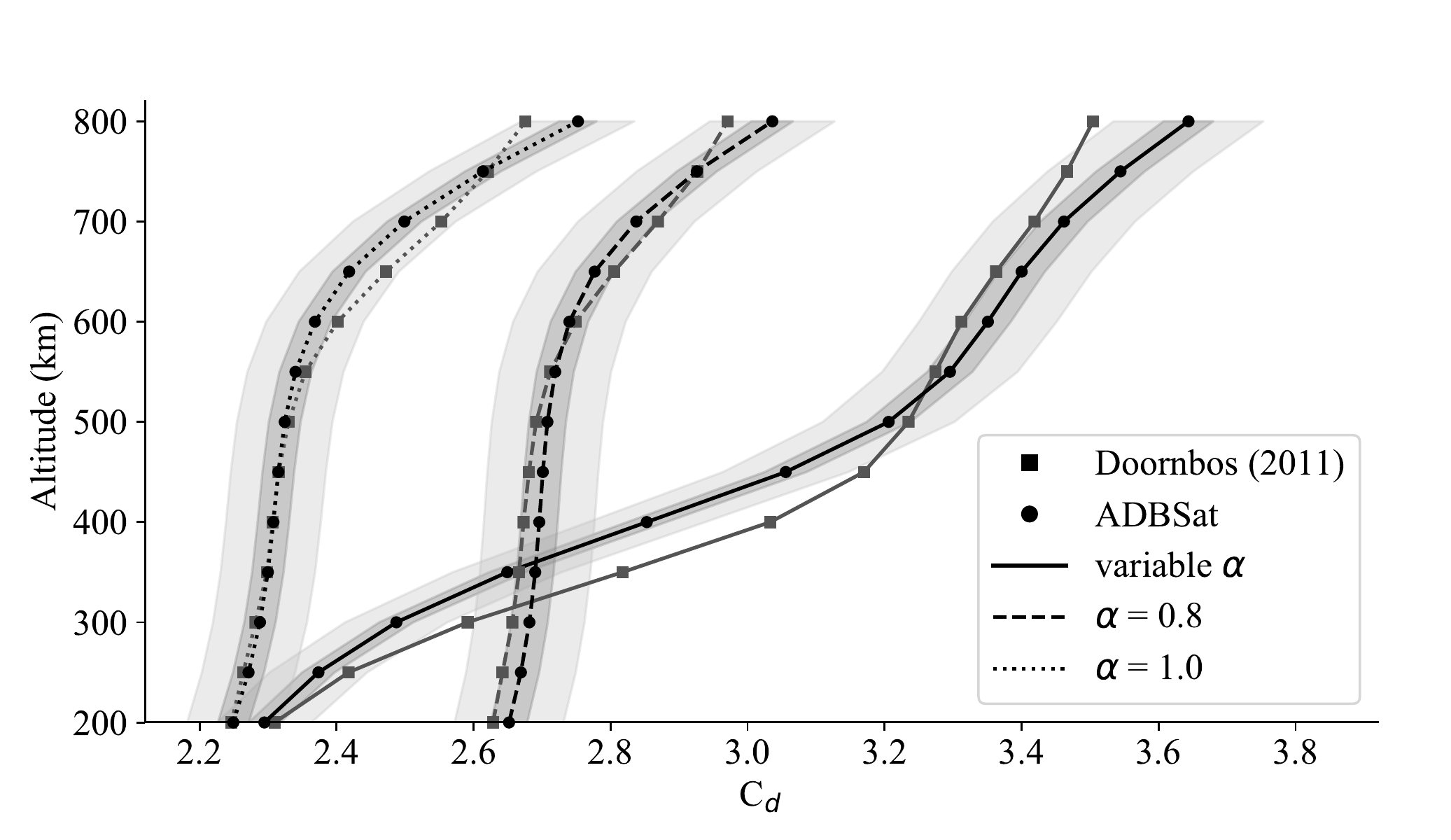}
         \caption{Solar minimum}
         \label{fig:CHAMP_solmin}
     \end{subfigure}
     
        \caption{Drag coefficients of CHAMP at a side-on view. 1\% and 3\% error ranges on ADBSat are highlighted. The variable $\alpha$ is calculated using the Langmuir isotherm model.}
        \label{fig:Doornbos}
\end{figure}

The analysis of CHAMP was complicated by the geometry available being slightly different to that which was used by \citet{EelcoThesis}. Comparison of the projected areas showed that the models were much more closely matched when viewed side-on, while the head-on projections did not align as closely. As the value of $C_d$ output by ADBSat relies heavily on the shape and its projected area, we could not directly compare our aerodynamic analysis in head-on flow conditions with that reported. 

\citet{EelcoThesis} presents $C_d$ for a number of scenarios where the satellite is side-on to the flow. We  tested the performance of ADBSat with a varying accommodation coefficient calculated from the Langmuir isotherm model \cite{AccommodationCoeffModel}, and fixed values of $\alpha = 1$ and $\alpha = 0.8$. We examined both maximum and minimum solar activity levels, presented in \cref{fig:CHAMP_solmax,fig:CHAMP_solmin} respectively. 

For the two cases of constant $\alpha$, some results agree as closely as 1\% to the published values, with all being consistently within 3\%. When we use a height-dependent $\alpha$, with all other parameters the same, a larger discrepancy is seen. However, it is of note that we calculated $\alpha$ independently, in an effort to reproduce the values used by \citet{EelcoThesis} - we do not know the values of $\alpha$ used therein. As $C_d$ is particularly sensitive to variations in $\alpha$, our conclusion is that the discrepancy is most likely due to this factor.

\section{Conclusions}
\label{sec:Conclusions}

By comparison to both DSMC and established literature sources, we have analysed the accuracy and limitations of ADBSat, a novel program that calculates the aerodynamics of a satellite body. While more simplistic than DSMC, the reduced computational and time cost of ADBSat is advantageous for some aspects of mission design. As an increasing number of satellite missions are developed to operate in VLEO, its efficiency will allow aerodynamic considerations to be employed earlier in the mission design process, and to explore a wider variety of designs. It can be used to quickly find promising satellite geometries that can then be analysed more thoroughly. It allows engineers lacking the extensive expertise required for DSMC simulations to obtain an approximate description of the satellite aerodynamics. Additionally, it can be used in a complementary manner to other aerodynamic analysis methods to obtain a thorough description of the aerodynamic characteristics of the test shape.

The comparison to DSMC, implemented through dsmcFoam, involved examination of specially designed test objects at a range of atmospheric conditions across VLEO altitudes. These objects included basic shapes, shapes that tested the new shading algorithm, and shapes that promoted multiple reflections of the atmospheric particles between the body panels. Analysis of the basic shapes showed a good agreement to both DSMC and closed-form models for strict FMF, where $K_n > 10$. The shading algorithm shows the desired performance at most incidence angles with respect to the flow. It breaks down if a large number of body panels are parallel to each other and to the flow, due to MATLAB's inherent floating point precision. Despite disregarding multiple particle reflections, ADBSat is shown to be accurate for some detailed shapes, but not for those where deep concavities (such as atmospheric intakes) are present on forward-facing sides of the body.

Comparison to existing literature detailing the drag analysis of real satellite shapes reveals that ADBSat differs by up to 3\% from the reported values. Therefore, it is recommended that an error interval of 3\% is adopted in the future for all ADBSat results. However, larger errors are seen when examining geometries that are unsuited to panel methods such as those with deep concavities in which particle trapping and multiple particle reflections can occur. It is recommended that alternative methods are used for the analysis of such cases.

\section*{Funding Sources}
The DISCOVERER project has received funding from the European Union’s Horizon 2020 research and innovation programme under grant agreement No 737183. Disclaimer: This publication reflects only the views of the authors. The European Commission is not liable for any use that may be made of the information contained therein.

\section*{Acknowledgments}
The authors would like to thank Marcin Pilinski and Eelco Doornbos for providing CAD satellite geometries to facilitate comparison to their work. We would also like to thank all members of the DISCOVERER project for their input. L. Sinpetru would like to thank her supervisors and colleagues for their useful comments.

\bibliographystyle{elsarticle-num-names}
\bibliography{sample.bib}

\begin{thebibliography}{64}
\expandafter\ifx\csname natexlab\endcsname\relax\def\natexlab#1{#1}\fi
\providecommand{\url}[1]{\texttt{#1}}
\providecommand{\href}[2]{#2}
\providecommand{\path}[1]{#1}
\providecommand{\DOIprefix}{doi:}
\providecommand{\ArXivprefix}{arXiv:}
\providecommand{\URLprefix}{URL: }
\providecommand{\Pubmedprefix}{pmid:}
\providecommand{\doi}[1]{\href{http://dx.doi.org/#1}{\path{#1}}}
\providecommand{\Pubmed}[1]{\href{pmid:#1}{\path{#1}}}
\providecommand{\bibinfo}[2]{#2}
\ifx\xfnm\relax \def\xfnm[#1]{\unskip,\space#1}\fi
\bibitem[{Roberts et~al.(2017)Roberts, Crisp, Edmondson, Haigh, Lyons, {Abrao
  Oiko}, {Macario Rojas}, Smith, Becedas, Gonz{\'a}lez, V{\'a}zquez, Bra{\~n}a,
  Antonini, Bay, Ghizoni, Jungnell, Morsb{\o}l, Binder, Boxberger, Herdrich,
  Romano, Fasoulas, Garcia-Almi{\~n}ana, Rodr{\'i}guez-Donaire, Kataria,
  Davidson, Outlaw, Belkouchi, Conte, Perez, Villain, and
  Schwalber}]{RobertsEtAl}
\bibinfo{author}{P.~Roberts}, \bibinfo{author}{N.~Crisp},
  \bibinfo{author}{S.~Edmondson}, \bibinfo{author}{S.~Haigh},
  \bibinfo{author}{R.~Lyons}, \bibinfo{author}{V.~{Abrao Oiko}},
  \bibinfo{author}{A.~{Macario Rojas}}, \bibinfo{author}{K.~Smith},
  \bibinfo{author}{J.~Becedas}, \bibinfo{author}{G.~Gonz{\'a}lez},
  \bibinfo{author}{I.~V{\'a}zquez}, \bibinfo{author}{{\'A}.~Bra{\~n}a},
  \bibinfo{author}{K.~Antonini}, \bibinfo{author}{K.~Bay},
  \bibinfo{author}{L.~Ghizoni}, \bibinfo{author}{V.~Jungnell},
  \bibinfo{author}{J.~Morsb{\o}l}, \bibinfo{author}{T.~Binder},
  \bibinfo{author}{A.~Boxberger}, \bibinfo{author}{G.~Herdrich},
  \bibinfo{author}{F.~Romano}, \bibinfo{author}{S.~Fasoulas},
  \bibinfo{author}{D.~Garcia-Almi{\~n}ana},
  \bibinfo{author}{S.~Rodr{\'i}guez-Donaire}, \bibinfo{author}{D.~Kataria},
  \bibinfo{author}{M.~Davidson}, \bibinfo{author}{R.~Outlaw},
  \bibinfo{author}{B.~Belkouchi}, \bibinfo{author}{A.~Conte},
  \bibinfo{author}{J.~Perez}, \bibinfo{author}{R.~Villain},
  \bibinfo{author}{A.~Schwalber},
\newblock \bibinfo{title}{{DISCOVERER}: Radical redesign of {Earth} observation
  satellites for sustained operation at significantly lower altitudes},
\newblock in: \bibinfo{booktitle}{68th International Astronautical Congress
  (IAC), Adelaide, Australia, 25-29 September}, \bibinfo{year}{2017}.
\bibitem[{{Virgili-Llop} et~al.(2014){Virgili-Llop}, Roberts, Hao, Ramio, and
  Beauplet}]{DefinitionOfVLEO}
\bibinfo{author}{J.~{Virgili-Llop}}, \bibinfo{author}{P.~Roberts},
  \bibinfo{author}{Z.~Hao}, \bibinfo{author}{L.~Ramio},
  \bibinfo{author}{V.~Beauplet},
\newblock \bibinfo{title}{Very low {Earth} orbit mission concepts for {Earth}
  observation. {Benefits} and challenges},
\newblock in: \bibinfo{booktitle}{Proceedings of the 12th Reinventing Space
  Conference, London, United Kingdom}, \bibinfo{year}{2014}.
\bibitem[{Reigber et~al.(2002)Reigber, L\"uhr, and
  Schwintzer}]{CHAMP_firstpaper}
\bibinfo{author}{C.~Reigber}, \bibinfo{author}{H.~L\"uhr},
  \bibinfo{author}{P.~Schwintzer},
\newblock \bibinfo{journal}{Adv. Space Res.}  30~(2) (\bibinfo{year}{2002})
  \bibinfo{pages}{129--134}.
  \bibinfo{note}{\url{https://doi.org/10.1016/S0273-1177(02)00276-4}}.
\bibitem[{Drinkwater et~al.(2006)Drinkwater, Haagmans, Muzi, Popescu,
  Floberghagen, Kern, and Fehringer}]{GOCE_firstpaper}
\bibinfo{author}{M.~R. Drinkwater}, \bibinfo{author}{R.~Haagmans},
  \bibinfo{author}{D.~Muzi}, \bibinfo{author}{A.~Popescu},
  \bibinfo{author}{R.~Floberghagen}, \bibinfo{author}{M.~Kern},
  \bibinfo{author}{M.~Fehringer},
\newblock \bibinfo{title}{The {GOCE} gravity mission: {ESA's} first core
  {Earth} explorer},
\newblock in: \bibinfo{booktitle}{Proceedings of the 3rd international {GOCE}
  user workshop, 6-8 November, 2006, Frescati, Italy}, \bibinfo{year}{2006},
  pp. \bibinfo{pages}{1--8}.
\bibitem[{Tapley et~al.(2004)Tapley, Bettadpur, Watkins, and
  Reigber}]{GRACE_firstpaper}
\bibinfo{author}{B.~D. Tapley}, \bibinfo{author}{S.~Bettadpur},
  \bibinfo{author}{M.~Watkins}, \bibinfo{author}{C.~Reigber},
\newblock \bibinfo{journal}{Geophys. Res. Lett.}  31~(9) (\bibinfo{year}{2004})
  \bibinfo{pages}{L09607}.
  \bibinfo{note}{\url{https://doi.org/10.1029/2004GL019920}}.
\bibitem[{Flechtner et~al.(2014)Flechtner, Morton, Watkins, and
  Webb}]{GRACE_FO}
\bibinfo{author}{F.~Flechtner}, \bibinfo{author}{P.~Morton},
  \bibinfo{author}{M.~Watkins}, \bibinfo{author}{F.~Webb},
\newblock \bibinfo{title}{Status of the {GRACE} follow-on mission},
\newblock in: \bibinfo{booktitle}{Gravity, Geoid and Height Systems},
  \bibinfo{publisher}{Springer International Publishing}, \bibinfo{year}{2014},
  pp. \bibinfo{pages}{117--121}.
\bibitem[{Fujita and Noda(2009)}]{SLATS}
\bibinfo{author}{K.~Fujita}, \bibinfo{author}{A.~Noda},
\newblock \bibinfo{title}{Rarefied aerodynamics of a super low altitude test
  satellite},
\newblock in: \bibinfo{booktitle}{41st AIAA Thermophysics Conference},
  \bibinfo{year}{2009}, p. \bibinfo{pages}{3606}.
  \bibinfo{note}{\url{https://doi.org/10.2514/6.2009-3606}}.
\bibitem[{Leomanni et~al.(2017)Leomanni, Garulli, Giannitrapani, and
  Scortecci}]{VLEOpropulsion}
\bibinfo{author}{M.~Leomanni}, \bibinfo{author}{A.~Garulli},
  \bibinfo{author}{A.~Giannitrapani}, \bibinfo{author}{F.~Scortecci},
\newblock \bibinfo{journal}{Acta Astronaut.}  133 (\bibinfo{year}{2017})
  \bibinfo{pages}{444 -- 454}.
  \bibinfo{note}{\url{https://doi.org/10.1016/j.actaastro.2016.11.001}}.
\bibitem[{Virgili-Llop et~al.(2019)Virgili-Llop, Polat, and
  Romano}]{VLEOstabilization}
\bibinfo{author}{J.~Virgili-Llop}, \bibinfo{author}{H.~C. Polat},
  \bibinfo{author}{M.~Romano},
\newblock \bibinfo{journal}{Front. Robot. AI}  6 (\bibinfo{year}{2019})
  \bibinfo{pages}{7}.
  \bibinfo{note}{\url{https://doi.org/10.3389/frobt.2019.00007}}.
\bibitem[{Andrews and Berthoud(2020)}]{VLEOionstuff}
\bibinfo{author}{S.~Andrews}, \bibinfo{author}{L.~Berthoud},
\newblock \bibinfo{journal}{Acta Astronaut.}  170 (\bibinfo{year}{2020})
  \bibinfo{pages}{386 -- 396}.
  \bibinfo{note}{\url{https://doi.org/10.1016/j.actaastro.2019.12.034}}.
\bibitem[{Crisp et~al.(2020)Crisp, Roberts, Livadiotti, Oiko, Edmondson, Haigh,
  Huyton, Sinpetru, Smith, Worrall, Becedas, Dom{\'i}nguez, Gonz{\'a}lez,
  Hanessian, M{\o}lgaard, Nielsen, Bisgaard, Chan, Fasoulas, Herdrich, Romano,
  Traub, Garc{\'i}a-Almi{\~n}ana, Rodr{\'i}guez-Donaire, Sureda, Kataria,
  Outlaw, Belkouchi, Conte, Perez, Villain, Hei{\ss}erer, and
  Schwalber}]{Crisp2020}
\bibinfo{author}{N.~Crisp}, \bibinfo{author}{P.~Roberts},
  \bibinfo{author}{S.~Livadiotti}, \bibinfo{author}{V.~Oiko},
  \bibinfo{author}{S.~Edmondson}, \bibinfo{author}{S.~Haigh},
  \bibinfo{author}{C.~Huyton}, \bibinfo{author}{L.~Sinpetru},
  \bibinfo{author}{K.~Smith}, \bibinfo{author}{S.~Worrall},
  \bibinfo{author}{J.~Becedas}, \bibinfo{author}{R.~Dom{\'i}nguez},
  \bibinfo{author}{D.~Gonz{\'a}lez}, \bibinfo{author}{V.~Hanessian},
  \bibinfo{author}{A.~M{\o}lgaard}, \bibinfo{author}{J.~Nielsen},
  \bibinfo{author}{M.~Bisgaard}, \bibinfo{author}{Y.-A. Chan},
  \bibinfo{author}{S.~Fasoulas}, \bibinfo{author}{G.~Herdrich},
  \bibinfo{author}{F.~Romano}, \bibinfo{author}{C.~Traub},
  \bibinfo{author}{D.~Garc{\'i}a-Almi{\~n}ana},
  \bibinfo{author}{S.~Rodr{\'i}guez-Donaire}, \bibinfo{author}{M.~Sureda},
  \bibinfo{author}{D.~Kataria}, \bibinfo{author}{R.~Outlaw},
  \bibinfo{author}{B.~Belkouchi}, \bibinfo{author}{A.~Conte},
  \bibinfo{author}{J.~Perez}, \bibinfo{author}{R.~Villain},
  \bibinfo{author}{B.~Hei{\ss}erer}, \bibinfo{author}{A.~Schwalber},
\newblock \bibinfo{journal}{Prog. Aerosp. Sci.}  117 (\bibinfo{year}{2020})
  \bibinfo{pages}{100619}.
  \bibinfo{note}{\url{https://doi.org/10.1016/j.paerosci.2020.100619}}.
\bibitem[{Sinpetru et~al.(????)Sinpetru, Crisp, Roberts, and
  Mostaza-Prieto}]{mySoftwarePaper}
\bibinfo{author}{L.~A. Sinpetru}, \bibinfo{author}{N.~H. Crisp},
  \bibinfo{author}{P.~C. Roberts}, \bibinfo{author}{D.~Mostaza-Prieto},
  \bibinfo{title}{{ADBSat}: Methodology of a novel panel method tool for
  aerodynamic analysis of satellites}. \bibinfo{note}{{Comput. Phys. Commun.
  (2022), in revision.}}
\bibitem[{Mostaza-Prieto(2017)}]{MostazaThesis}
\bibinfo{author}{D.~Mostaza-Prieto}, \bibinfo{title}{Characterisation and
  Applications of Aerodynamic Torques on Satellites}, Ph.D. thesis, The
  University of Manchester, \bibinfo{year}{2017}.
\bibitem[{Livadiotti et~al.(2020)Livadiotti, Crisp, Roberts, Worrall, Oiko,
  Edmondson, Haigh, Huyton, Smith, Sinpetru, Holmes, Becedas, Domínguez,
  Cañas, Christensen, Mølgaard, Nielsen, Bisgaard, Chan, Herdrich, Romano,
  Fasoulas, Traub, Garcia-Almiñana, Rodr{\'i}guez-Donaire, Sureda, Kataria,
  Belkouchi, Conte, Perez, Villain, and Outlaw}]{sabrina}
\bibinfo{author}{S.~Livadiotti}, \bibinfo{author}{N.~H. Crisp},
  \bibinfo{author}{P.~C. Roberts}, \bibinfo{author}{S.~D. Worrall},
  \bibinfo{author}{V.~T. Oiko}, \bibinfo{author}{S.~Edmondson},
  \bibinfo{author}{S.~J. Haigh}, \bibinfo{author}{C.~Huyton},
  \bibinfo{author}{K.~L. Smith}, \bibinfo{author}{L.~A. Sinpetru},
  \bibinfo{author}{B.~E. Holmes}, \bibinfo{author}{J.~Becedas},
  \bibinfo{author}{R.~M. Domínguez}, \bibinfo{author}{V.~Cañas},
  \bibinfo{author}{S.~Christensen}, \bibinfo{author}{A.~Mølgaard},
  \bibinfo{author}{J.~Nielsen}, \bibinfo{author}{M.~Bisgaard},
  \bibinfo{author}{Y.-A. Chan}, \bibinfo{author}{G.~H. Herdrich},
  \bibinfo{author}{F.~Romano}, \bibinfo{author}{S.~Fasoulas},
  \bibinfo{author}{C.~Traub}, \bibinfo{author}{D.~Garcia-Almiñana},
  \bibinfo{author}{S.~Rodr{\'i}guez-Donaire}, \bibinfo{author}{M.~Sureda},
  \bibinfo{author}{D.~Kataria}, \bibinfo{author}{B.~Belkouchi},
  \bibinfo{author}{A.~Conte}, \bibinfo{author}{J.~S. Perez},
  \bibinfo{author}{R.~Villain}, \bibinfo{author}{R.~Outlaw},
\newblock \bibinfo{journal}{Prog. Aerosp. Sci.}  119 (\bibinfo{year}{2020})
  \bibinfo{pages}{100675}.
  \bibinfo{note}{\url{https://doi.org/10.1016/j.paerosci.2020.100675}}.
\bibitem[{Moss et~al.(2006{\natexlab{a}})Moss, Boyles, and Greene}]{orion}
\bibinfo{author}{J.~Moss}, \bibinfo{author}{K.~Boyles},
  \bibinfo{author}{F.~Greene},
\newblock \bibinfo{title}{Orion aerodynamics for hypersonic free molecular to
  continuum conditions},
\newblock in: \bibinfo{booktitle}{14th AIAA{/}AHI Space Planes and Hypersonic
  Systems and Technologies Conference}, \bibinfo{year}{2006}{\natexlab{a}}.
  \bibinfo{note}{\url{https://doi.org/10.2514/6.2006-8081}}.
\bibitem[{Moss et~al.(2006{\natexlab{b}})Moss, Glass, and Greene}]{apollo}
\bibinfo{author}{J.~Moss}, \bibinfo{author}{C.~Glass},
  \bibinfo{author}{F.~Greene},
\newblock \bibinfo{title}{{DSMC} simulations of {Apollo} capsule aerodynamics
  for hypersonic rarefied conditions},
\newblock in: \bibinfo{booktitle}{9th AIAA/ASME Joint Thermophysics and Heat
  Transfer Conference}, \bibinfo{year}{2006}{\natexlab{b}}, p.
  \bibinfo{pages}{3577}.
  \bibinfo{note}{\url{https://doi.org/10.2514/6.2006-3577}}.
\bibitem[{Fuller and Tolson(2009)}]{FreeMat}
\bibinfo{author}{J.~D. Fuller}, \bibinfo{author}{R.~H. Tolson},
\newblock \bibinfo{journal}{J. Spacecraft Rockets}  46~(5)
  (\bibinfo{year}{2009}) \bibinfo{pages}{938--948}.
  \bibinfo{note}{\url{https://doi.org/10.2514/1.43205}}.
\bibitem[{Fredo and Kaplan(1981)}]{FreeMac}
\bibinfo{author}{R.~M. Fredo}, \bibinfo{author}{M.~H. Kaplan},
\newblock \bibinfo{journal}{J. Spacecraft Rockets}  18~(4)
  (\bibinfo{year}{1981}) \bibinfo{pages}{367--373}.
  \bibinfo{note}{\url{https://doi.org/10.2514/3.28061}}.
\bibitem[{Bird(1994)}]{bird1994}
\bibinfo{author}{G.~Bird}, \bibinfo{title}{Molecular Gas Dynamics and the
  Direct Simulation of Gas Flows}, The Oxford engineering science series,
  \bibinfo{publisher}{Clarendon Press}, \bibinfo{year}{1994}.
\bibitem[{Mostaza-Prieto et~al.(2014)Mostaza-Prieto, Graziano, and
  Roberts}]{DragModelling}
\bibinfo{author}{D.~Mostaza-Prieto}, \bibinfo{author}{B.~P. Graziano},
  \bibinfo{author}{P.~C.~E. Roberts},
\newblock \bibinfo{journal}{Prog. Aerosp. Sci.}  64 (\bibinfo{year}{2014})
  \bibinfo{pages}{56--65}.
  \bibinfo{note}{\url{https://doi.org/10.1016/j.paerosci.2013.09.001}}.
\bibitem[{White et~al.(2018)White, Borg, Scanlon, Longshaw, John, Emerson, and
  Reese}]{dsmcFoamPlus}
\bibinfo{author}{C.~White}, \bibinfo{author}{M.~K. Borg},
  \bibinfo{author}{T.~J. Scanlon}, \bibinfo{author}{S.~M. Longshaw},
  \bibinfo{author}{B.~John}, \bibinfo{author}{D.~R. Emerson},
  \bibinfo{author}{J.~M. Reese},
\newblock \bibinfo{journal}{Comput. Phys. Commun.}  224 (\bibinfo{year}{2018})
  \bibinfo{pages}{22 -- 43}.
  \bibinfo{note}{\url{https://doi.org/10.1016/j.cpc.2017.09.030}}.
\bibitem[{Palharini et~al.(2015)Palharini, White, Scanlon, Brown, Borg, and
  Reese}]{dsmcFoam4}
\bibinfo{author}{R.~C. Palharini}, \bibinfo{author}{C.~White},
  \bibinfo{author}{T.~J. Scanlon}, \bibinfo{author}{R.~E. Brown},
  \bibinfo{author}{M.~K. Borg}, \bibinfo{author}{J.~M. Reese},
\newblock \bibinfo{journal}{Comput. Fluids}  120 (\bibinfo{year}{2015})
  \bibinfo{pages}{140 -- 157}.
  \bibinfo{note}{\url{https://doi.org/10.1016/j.compfluid.2015.07.021}}.
\bibitem[{White et~al.(2013)White, Colombo, Scanlon, McInnes, and
  Reese}]{dsmcFoam2}
\bibinfo{author}{C.~White}, \bibinfo{author}{C.~Colombo},
  \bibinfo{author}{T.~J. Scanlon}, \bibinfo{author}{C.~R. McInnes},
  \bibinfo{author}{J.~M. Reese},
\newblock \bibinfo{journal}{Adv. Space Res.}  51~(11) (\bibinfo{year}{2013})
  \bibinfo{pages}{2112 -- 2124}.
  \bibinfo{note}{\url{https://doi.org/10.1016/j.asr.2013.01.002}}.
\bibitem[{Park et~al.(2014)Park, Myong, Kim, and Baek}]{ParkEtAlShapeOpt}
\bibinfo{author}{J.~H. Park}, \bibinfo{author}{R.~S. Myong},
  \bibinfo{author}{D.~H. Kim}, \bibinfo{author}{S.~W. Baek},
\newblock \bibinfo{journal}{AIP Conf. Proc.}  1628~(1) (\bibinfo{year}{2014})
  \bibinfo{pages}{1331--1336}.
  \bibinfo{note}{\url{https://doi.org/10.1063/1.4902745}}.
\bibitem[{Parodi et~al.(2009)Parodi, Le~Quang, Bariselli, Boccelli, Alsalihi,
  and Magin}]{intakeDesign}
\bibinfo{author}{P.~Parodi}, \bibinfo{author}{D.~Le~Quang},
  \bibinfo{author}{F.~Bariselli}, \bibinfo{author}{S.~Boccelli},
  \bibinfo{author}{Z.~Alsalihi}, \bibinfo{author}{T.~Magin},
\newblock \bibinfo{title}{Study of a collector-intake system for {VLEO}
  air-breathing platforms},
\newblock in: \bibinfo{booktitle}{International Conference on Flight vehicles,
  Aerothermodynamics and Re-entry Missions and Engineering (FAR), Monopoli,
  Italy}, \bibinfo{year}{2009}.
\bibitem[{Mehta et~al.(2014)Mehta, Walker, McLaughlin, and
  Koller}]{ComparingDragCoeffs_GSIs}
\bibinfo{author}{P.~M. Mehta}, \bibinfo{author}{A.~Walker},
  \bibinfo{author}{C.~A. McLaughlin}, \bibinfo{author}{J.~Koller},
\newblock \bibinfo{journal}{J. Spacecraft Rockets}  51~(3)
  (\bibinfo{year}{2014}) \bibinfo{pages}{873--883}.
  \bibinfo{note}{\url{https://doi.org/10.2514/1.A32566}}.
\bibitem[{Pilinski et~al.(2011)Pilinski, Argrow, and
  Palo}]{DSMCworkaroundDevised}
\bibinfo{author}{M.~D. Pilinski}, \bibinfo{author}{B.~M. Argrow},
  \bibinfo{author}{S.~E. Palo},
\newblock \bibinfo{journal}{J. Spacecraft Rockets}  48~(2)
  (\bibinfo{year}{2011}) \bibinfo{pages}{312--325}.
  \bibinfo{note}{\url{https://doi.org/10.2514/1.50915}}.
\bibitem[{Scanlon et~al.(2010)Scanlon, Roohi, White, Darbandi, and
  Reese}]{dsmcFoam}
\bibinfo{author}{T.~J. Scanlon}, \bibinfo{author}{E.~Roohi},
  \bibinfo{author}{C.~White}, \bibinfo{author}{M.~Darbandi},
  \bibinfo{author}{J.~M. Reese},
\newblock \bibinfo{journal}{Comput. Fluids}  39~(10) (\bibinfo{year}{2010})
  \bibinfo{pages}{2078 -- 2089}.
  \bibinfo{note}{\url{https://doi.org/10.1016/j.compfluid.2010.07.014}}.
\bibitem[{Moe and Moe(2005)}]{MoeMoe2005}
\bibinfo{author}{K.~Moe}, \bibinfo{author}{M.~M. Moe},
\newblock \bibinfo{journal}{Planet. Space Sci.}  53~(8) (\bibinfo{year}{2005})
  \bibinfo{pages}{793 -- 801}.
  \bibinfo{note}{\url{https://doi.org/10.1016/j.pss.2005.03.005}}.
\bibitem[{Weller et~al.(1998)Weller, Tabor, Jasak, and Fureby}]{OFcitation}
\bibinfo{author}{H.~G. Weller}, \bibinfo{author}{G.~Tabor},
  \bibinfo{author}{H.~Jasak}, \bibinfo{author}{C.~Fureby},
\newblock \bibinfo{journal}{Comput. Phys.}  12~(6) (\bibinfo{year}{1998})
  \bibinfo{pages}{620--631}.
  \bibinfo{note}{\url{https://doi.org/10.1063/1.168744}}.
\bibitem[{Weinberg(2014)}]{DSMCcriteria}
\bibinfo{author}{M.~D. Weinberg},
\newblock \bibinfo{journal}{Mon. Not. R. Astron. Soc.}  438~(4)
  (\bibinfo{year}{2014}) \bibinfo{pages}{2995--3006}.
  \bibinfo{note}{\url{https://doi.org/10.1093/mnras/stt2406}}.
\bibitem[{Bird(2007)}]{bird2007dsmc}
\bibinfo{author}{G.~A. Bird},
\newblock \bibinfo{title}{Sophisticated {DSMC}},
\newblock in: \bibinfo{booktitle}{Notes prepared for a short course at the
  DSMC07 meeting, Santa Fe, USA}, \bibinfo{year}{2007}.
\bibitem[{Zuppardi et~al.(2015)Zuppardi, Morsa, Savino, Sippel, and
  Schwanekamp}]{MaxwellVsCLL}
\bibinfo{author}{G.~Zuppardi}, \bibinfo{author}{L.~Morsa},
  \bibinfo{author}{R.~Savino}, \bibinfo{author}{M.~Sippel},
  \bibinfo{author}{T.~Schwanekamp},
\newblock \bibinfo{journal}{Eur. J. Mech. B-Fluid.}  53 (\bibinfo{year}{2015})
  \bibinfo{pages}{37 -- 47}.
  \bibinfo{note}{\url{https://doi.org/10.1016/j.euromechflu.2015.04.003}}.
\bibitem[{Padilla and Boyd(2009)}]{MaxwellVsCLL2}
\bibinfo{author}{J.~Padilla}, \bibinfo{author}{I.~Boyd},
\newblock \bibinfo{journal}{J. Thermophys. Heat Tr.}  23~(1)
  (\bibinfo{year}{2009}) \bibinfo{pages}{96--105}.
  \bibinfo{note}{\url{https://doi.org/10.2514/1.36375}}.
\bibitem[{Wadsworth et~al.(2003)Wadsworth, {VanGilder}, and Dogra}]{GSIforDSMC}
\bibinfo{author}{D.~C. Wadsworth}, \bibinfo{author}{D.~B. {VanGilder}},
  \bibinfo{author}{V.~K. Dogra},
\newblock \bibinfo{journal}{AIP Conf. Proc.}  663~(1) (\bibinfo{year}{2003})
  \bibinfo{pages}{965--972}.
  \bibinfo{note}{\url{https://doi.org/10.1063/1.1581644}}.
\bibitem[{Sentman(1961)}]{sentman}
\bibinfo{author}{L.~H. Sentman}, \bibinfo{title}{Free molecular flow theory and
  its application to the determination of aerodynamic forces},
  \bibinfo{type}{Technical Report} \bibinfo{number}{LMSC-448514}, Lockheed
  Missiles and Space Co., \bibinfo{address}{Sunnyvale, CA},
  \bibinfo{year}{1961}.
\bibitem[{Schaaf and Chambre(1961)}]{SchaafCham}
\bibinfo{author}{S.~A. Schaaf}, \bibinfo{author}{P.~L. Chambre},
  \bibinfo{title}{Flow of Rarefied Gases}, volume~\bibinfo{volume}{8},
  \bibinfo{publisher}{Princeton University Press}, \bibinfo{address}{Princeton,
  NJ}, \bibinfo{year}{1961}.
\bibitem[{Cook(1965)}]{Cook}
\bibinfo{author}{G.~Cook},
\newblock \bibinfo{journal}{Planet. Space Sci.}  13~(10) (\bibinfo{year}{1965})
  \bibinfo{pages}{929 -- 946}.
  \bibinfo{note}{\url{https://doi.org/10.1016/0032-0633(65)90150-9}}.
\bibitem[{Cercignani and Lampis(1971)}]{CLLmodel}
\bibinfo{author}{C.~Cercignani}, \bibinfo{author}{M.~Lampis},
\newblock \bibinfo{journal}{Transport Theor. Stat.}  1~(2)
  (\bibinfo{year}{1971}) \bibinfo{pages}{101--114}.
  \bibinfo{note}{\url{https://doi.org/10.1080/00411457108231440}}.
\bibitem[{Lord(1991)}]{CLL_Lord}
\bibinfo{author}{R.~G. Lord},
\newblock \bibinfo{journal}{Phys.Fluids A-Fluid}  3~(4) (\bibinfo{year}{1991})
  \bibinfo{pages}{706--710}.
  \bibinfo{note}{\url{https://doi.org/10.1063/1.858076}}.
\bibitem[{Storch(2002)}]{storchHyp}
\bibinfo{author}{J.~A. Storch}, \bibinfo{title}{Aerodynamic disturbances on
  spacecraft in free-molecular flow (monograph)}, \bibinfo{type}{Technical
  Report}, The Aerospace Corporation, \bibinfo{year}{2002}.
\bibitem[{Moe et~al.(1998)Moe, Moe, and Wallace}]{MoeMoe1998}
\bibinfo{author}{K.~Moe}, \bibinfo{author}{M.~M. Moe}, \bibinfo{author}{S.~D.
  Wallace},
\newblock \bibinfo{journal}{J. Spacecraft Rockets}  35~(3)
  (\bibinfo{year}{1998}) \bibinfo{pages}{266--272}.
  \bibinfo{note}{\url{https://doi.org/10.2514/2.3350}}.
\bibitem[{Pardini et~al.(2010)Pardini, Anselmo, Moe, and
  Moe}]{AccommodationCoeffPardini}
\bibinfo{author}{C.~Pardini}, \bibinfo{author}{L.~Anselmo},
  \bibinfo{author}{K.~Moe}, \bibinfo{author}{M.~Moe},
\newblock \bibinfo{journal}{Adv. Space Res.}  45~(5) (\bibinfo{year}{2010})
  \bibinfo{pages}{638 -- 650}.
  \bibinfo{note}{\url{https://doi.org/10.1016/j.asr.2009.08.034}}.
\bibitem[{Pilinski et~al.(2010)Pilinski, Argrow, and
  Palo}]{AccommodationCoeffModel}
\bibinfo{author}{M.~D. Pilinski}, \bibinfo{author}{B.~M. Argrow},
  \bibinfo{author}{S.~E. Palo},
\newblock \bibinfo{journal}{J. Spacecraft Rockets}  47~(6)
  (\bibinfo{year}{2010}) \bibinfo{pages}{951--956}.
  \bibinfo{note}{\url{https://doi.org/10.2514/1.49330}}.
\bibitem[{Mehta et~al.(2013)Mehta, McLaughlin, and
  Sutton}]{DSMCworkaroundImplemented}
\bibinfo{author}{P.~M. Mehta}, \bibinfo{author}{C.~A. McLaughlin},
  \bibinfo{author}{E.~K. Sutton},
\newblock \bibinfo{journal}{Adv. Space Res.}  52~(12) (\bibinfo{year}{2013})
  \bibinfo{pages}{2035 -- 2051}.
  \bibinfo{note}{\url{https://doi.org/10.1016/j.asr.2013.08.033}}.
\bibitem[{Moe et~al.(1995)Moe, Wallace, and Moe}]{MoeMoe1995}
\bibinfo{author}{M.~M. Moe}, \bibinfo{author}{S.~D. Wallace},
  \bibinfo{author}{K.~Moe}, \bibinfo{title}{in: R.M. Johnson, T.L. Killeen
  (Eds.) The Upper Mesosphere and Lower Thermosphere: A Review of Experiment
  and Theory}, \bibinfo{publisher}{American Geophysical Union (AGU)},
  \bibinfo{address}{Washington, D. C.}, \bibinfo{year}{1995}, pp.
  \bibinfo{pages}{349--356}.
\bibitem[{{Yi\v{g}it} and Medvedev(2010)}]{F10.7_definitions}
\bibinfo{author}{E.~{Yi\v{g}it}}, \bibinfo{author}{A.~S. Medvedev},
\newblock \bibinfo{journal}{J. Geophys. Res. Space Phys.}  115
  (\bibinfo{year}{2010}) \bibinfo{pages}{A8}.
  \bibinfo{note}{\url{https://doi.org/10.1029/2009JA015106}}.
\bibitem[{Picone et~al.(2002)Picone, Hedin, Drob, and Aikin}]{NRLMSISE00}
\bibinfo{author}{J.~M. Picone}, \bibinfo{author}{A.~E. Hedin},
  \bibinfo{author}{D.~P. Drob}, \bibinfo{author}{A.~C. Aikin},
\newblock \bibinfo{journal}{J. Geophys. Res. Space Phys.}  107~(A12)
  (\bibinfo{year}{2002}) \bibinfo{pages}{15--16}.
  \bibinfo{note}{\url{https://doi.org/10.1029/2002JA009430}}.
\bibitem[{Bowman et~al.(2008)Bowman, Tobiska, Marcos, and Valladares}]{JB2006}
\bibinfo{author}{B.~R. Bowman}, \bibinfo{author}{W.~K. Tobiska},
  \bibinfo{author}{F.~A. Marcos}, \bibinfo{author}{C.~Valladares},
\newblock \bibinfo{journal}{J. Atmos. Sol.-Terr. Phys.}  70~(5)
  (\bibinfo{year}{2008}) \bibinfo{pages}{774 -- 793}.
  \bibinfo{note}{\url{https://doi.org/10.1016/j.jastp.2007.10.002}}.
\bibitem[{Bowman et~al.(2009)Bowman, Tobiska, Marcos, Huang, Lin, and
  Burke}]{JB2008}
\bibinfo{author}{B.~Bowman}, \bibinfo{author}{W.~K. Tobiska},
  \bibinfo{author}{F.~Marcos}, \bibinfo{author}{C.~Huang},
  \bibinfo{author}{C.~Lin}, \bibinfo{author}{W.~Burke},
\newblock \bibinfo{title}{A new empirical thermospheric density model {JB2008}
  using new solar and geomagnetic indices},
\newblock in: \bibinfo{booktitle}{37th COSPAR Scientific Assembly},
  volume~\bibinfo{volume}{37} of \textit{\bibinfo{series}{COSPAR Meeting}},
  \bibinfo{year}{2009}, p. \bibinfo{pages}{367}.
  \bibinfo{note}{\url{https://doi.org/10.2514/6.2008-6438}}.
\bibitem[{Bruinsma et~al.(2003)Bruinsma, Thuillier, and Barlier}]{DTM2000}
\bibinfo{author}{S.~Bruinsma}, \bibinfo{author}{G.~Thuillier},
  \bibinfo{author}{F.~Barlier},
\newblock \bibinfo{journal}{J. Atmos. Sol.-Terr. Phys.}  65~(9)
  (\bibinfo{year}{2003}) \bibinfo{pages}{1053 -- 1070}.
  \bibinfo{note}{\url{https://doi.org/10.1016/S1364-6826(03)00137-8}}.
\bibitem[{{Bruinsma}(2015)}]{DTM2013}
\bibinfo{author}{S.~{Bruinsma}},
\newblock \bibinfo{journal}{J. Space Weather Spac.}  5 (\bibinfo{year}{2015})
  \bibinfo{pages}{A1}.
  \bibinfo{note}{\url{https://doi.org/10.1051/swsc/2015001}}.
\bibitem[{Zhaborovskyi(2014)}]{JB_NRLMSISE_DTM_comparison}
\bibinfo{author}{V.~P. Zhaborovskyi},
\newblock \bibinfo{journal}{Kinemat. Phys. Celest. Bodies}  30~(6)
  (\bibinfo{year}{2014}) \bibinfo{pages}{308--312}.
  \bibinfo{note}{\url{https://doi.org/10.3103/S0884591314060087}}.
\bibitem[{{Bruinsma} et~al.(2012){Bruinsma}, {S\'anchez-Ortiz}, {Olmedo}, and
  {Guijarro}}]{DTM2009}
\bibinfo{author}{S.~L. {Bruinsma}}, \bibinfo{author}{N.~{S\'anchez-Ortiz}},
  \bibinfo{author}{E.~{Olmedo}}, \bibinfo{author}{N.~{Guijarro}},
\newblock \bibinfo{journal}{J. Space Weather Spac.}  2 (\bibinfo{year}{2012})
  \bibinfo{pages}{A04}.
  \bibinfo{note}{\url{https://doi.org/10.1051/swsc/2012005}}.
\bibitem[{Siskind et~al.(2004)Siskind, Picone, Stevens, and
  Minschwaner}]{NRLMSISE_rocket_data}
\bibinfo{author}{D.~E. Siskind}, \bibinfo{author}{J.~M. Picone},
  \bibinfo{author}{M.~H. Stevens}, \bibinfo{author}{K.~Minschwaner},
\newblock \bibinfo{journal}{J. Geophys. Res. Space Phys.}  109~(A1)
  (\bibinfo{year}{2004}) \bibinfo{pages}{A01303}.
  \bibinfo{note}{\url{https://doi.org/10.1029/2003JA009943}}.
\bibitem[{Doornbos(2011)}]{EelcoThesis}
\bibinfo{author}{E.~N. Doornbos}, \bibinfo{title}{Thermospheric Density and
  Wind Determination from Satellite Dynamics}, Ph.D. thesis, Delft University
  of Technology, \bibinfo{year}{2011}.
\bibitem[{Zhou et~al.(2009)Zhou, Ma, Lühr, Xiong, and
  Reigber}]{NRLMSISE_champ_correction}
\bibinfo{author}{Y.~Zhou}, \bibinfo{author}{S.~Ma}, \bibinfo{author}{H.~Lühr},
  \bibinfo{author}{C.~Xiong}, \bibinfo{author}{C.~Reigber},
\newblock \bibinfo{journal}{Adv. Space Res.}  43~(5) (\bibinfo{year}{2009})
  \bibinfo{pages}{819 -- 828}.
  \bibinfo{note}{\url{https://doi.org/10.1016/j.asr.2008.06.016}}.
\bibitem[{{Drob} et~al.(2018){Drob}, {Emmert}, {Siskind}, and
  {Picone}}]{NRLMSISE_2}
\bibinfo{author}{D.~{Drob}}, \bibinfo{author}{J.~{Emmert}},
  \bibinfo{author}{D.~{Siskind}}, \bibinfo{author}{J.~M. {Picone}},
\newblock \bibinfo{title}{{NRLMSIS 2.0: New Formulation, New Data}},
\newblock in: \bibinfo{booktitle}{42nd COSPAR Scientific Assembly},
  volume~\bibinfo{volume}{42}, \bibinfo{year}{2018}, pp.
  \bibinfo{pages}{C4.2--3--18}.
\bibitem[{Cheng et~al.(2019)Cheng, Yang, Xiao, and Hu}]{cheng2019density}
\bibinfo{author}{X.~Cheng}, \bibinfo{author}{J.~Yang},
  \bibinfo{author}{C.~Xiao}, \bibinfo{author}{X.~Hu},
\newblock \bibinfo{journal}{Ann. Geophys. Discussions}  2019
  (\bibinfo{year}{2019}) \bibinfo{pages}{1--18}.
  \bibinfo{note}{\url{https://doi.org/10.5194/angeo-2019-93}}.
\bibitem[{Dai et~al.(2020)Dai, Pan, Hu, Bai, Ban, Zhang, and
  Che}]{2019NNimprovement}
\bibinfo{author}{Y.~Dai}, \bibinfo{author}{W.~Pan}, \bibinfo{author}{X.~Hu},
  \bibinfo{author}{Z.~Bai}, \bibinfo{author}{C.~Ban},
  \bibinfo{author}{H.~Zhang}, \bibinfo{author}{Y.~Che},
\newblock \bibinfo{journal}{Meteorol. Atmos. Phys.}  132 (\bibinfo{year}{2020})
  \bibinfo{pages}{451--459}.
  \bibinfo{note}{\url{https://doi.org/10.1007/s00703-019-00700-w}}.
\bibitem[{Mehta et~al.(2014)Mehta, Walker, Lawrence, Linares, Higdon, and
  Koller}]{responseSurfacesMehta}
\bibinfo{author}{P.~M. Mehta}, \bibinfo{author}{A.~Walker},
  \bibinfo{author}{E.~Lawrence}, \bibinfo{author}{R.~Linares},
  \bibinfo{author}{D.~Higdon}, \bibinfo{author}{J.~Koller},
\newblock \bibinfo{journal}{Adv. Space Res.}  54~(8) (\bibinfo{year}{2014})
  \bibinfo{pages}{1590 -- 1607}.
  \bibinfo{note}{\url{https://doi.org/10.1016/j.asr.2014.06.033}}.
\bibitem[{March et~al.(2019)March, Doornbos, and Visser}]{HighFidelityModels}
\bibinfo{author}{G.~March}, \bibinfo{author}{E.~Doornbos},
  \bibinfo{author}{P.~Visser},
\newblock \bibinfo{journal}{Adv. Space Res.}  63~(1) (\bibinfo{year}{2019})
  \bibinfo{pages}{213 -- 238}.
  \bibinfo{note}{\url{https://doi.org/10.1016/j.asr.2018.07.009}}.
\bibitem[{Sutton(2008)}]{SuttonThesis}
\bibinfo{author}{E.~K. Sutton}, \bibinfo{title}{Effects of solar disturbances
  on the thermosphere densities and winds from CHAMP and GRACE satellite
  accelerometer data}, Ph.D. thesis, University of Colorado Boulder,
  \bibinfo{year}{2008}.
\bibitem[{Bowman et~al.(2008)Bowman, Marcos, Moe, and Moe}]{bowmanGRACECHAMP}
\bibinfo{author}{B.~R. Bowman}, \bibinfo{author}{F.~Marcos},
  \bibinfo{author}{K.~Moe}, \bibinfo{author}{M.~Moe},
\newblock \bibinfo{journal}{Adv. Astronaut. Sci.}  129~(1)
  (\bibinfo{year}{2008}) \bibinfo{pages}{147--166}.

\end{thebibliography}

\end{document}